\documentclass[12pt,preprint]{aastex} 

\newcommand{\skipthis}[1]{} 
\newcommand{\hii}{{\rm H}{\sc ii}}
\newcommand{\uchii}{{\rm UCH}{\sc ii}} 
\def\nh3{$\rm{NH_3}$}
\def\NH3{$\rm{NH_3}$} 
 
\def\msun{M$_\odot$}
 
\def\lsun{L$_\odot$} 
\def\kms-1{km~s$^{-1}$}
\def\h2o{$\rm{H_2O}$} 
\def\h2{$\rm{H_2}$} 
\def\CM2{$\rm{cm^{-2}}$}
\def\cm3{$\rm{cm^{-3}}$} 
 
\def\vlsr{$\rm{V_{LSR}}$}

\newcommand{\lsim}{${\raisebox{-.9ex}{$\stackrel{\textstyle<}{\sim}$}}$ }
\newcommand{\gsim}{${\raisebox{-.9ex}{$\stackrel{\textstyle>}{\sim}$}}$ }

\begin{document}


\title {Multiple Jets from the High-Mass (Proto)stellar Cluster AFGL5142}

\author{Qizhou Zhang$^1$, Todd R.  Hunter$^1$, H.  Beuther$^2$, T.  K.
Sridharan$^1$, S.-Y.  Liu$^3$, Y.-N.  Su$^3$, H.-R.  Chen$^3$, Y. Chen$^1$}
\affil{$^1$Harvard Smithsonian Center for Astrophysics\\ 60 Garden Street\\
Cambridge, Massachusetts 02138, USA}

\affil{$^2$Max-Planck-Institute for Astronomy \\
K\"onigstuhl 17, 69117 Heidelberg, Germany}

\affil{$^3$Academia Sinica Institute of Astronomy and Astrophysics, P.O.  Box
23-141, Taipei 106, Taiwan}

\begin{abstract} 
We present studies of a massive protocluster AFGL5142 in
the J=2-1 transition of the CO isotopologues, SO, CH$_3$OH and CH$_3$CN lines, as
well as continuum at 225 GHz and 8.4 GHz.  The 225 GHz continuum emission reveals
three prominent peaks MM-1, MM-2 and MM-3 with estimated circumstellar
material of 3, 3,
and 2 \msun, respectively.  MM-1 and MM-2 are associated with strong CH$_3$CN
emission with temperatures of $90 \pm 20$ and $250 \pm 40$ K, respectively, while
both MM-1 and MM-3 are associated with faint continuum 
emission at 8.4 GHz. The heating implied by the temperature
indicates that
MM-1 and MM-2 cores contain embedded massive young stars. Additional dust
continuum peaks MM-4 and MM-5 appear to be associated with \h2O masers.
With many continuum sources at cm and mm wavelengths, and those already
identified in the infrared, this region is forming a cluster of stars.

A total of 22 lines from 9 molecules are detected. The line strength
varies remarkably in the region. The strong SO emission is
found both in molecular outflows and cloud cores. CH$_3$OH emission, on the
contrary, is much weaker in molecular outflows, and is
detected toward hot cores MM-1 and MM-2 only, 
but is absent in less massive and perhaps less evolved cores
MM-3, MM-4 and MM-5. The modeling of the CH$_3$CN spectra yields
an abundance relative to \h2
of $1 \times 10^{-8}$ and $4 \times 10^{-8}$ for the MM-1 and MM-2
cores, respectively. With similar
core mass, the higher temperature and CH$_3$CN abundance in the
MM-2 core suggest that it might be at a more evolved stage than the MM-1
core.

The CO and SO emission reveals at least three molecular outflows originating
from the center of the dust core.  The outflows are well collimated, with terminal
velocities up to 50 \kms-1\  from the cloud velocity.  Outflow A
coincides with the SiO jet identified previously by \citet{hunter1999}.
The maximum velocity of both the CO and SiO outflow increases with the
distance from the driving source.  Since
jet-like outflows and disk-mediated 
accretion process are physically connected,  the well
collimated outflows indicate that even in this cluster environment, accretion
is responsible for the formation of individual stars in the cluster.

\end{abstract}


\keywords{ISM:  kinematics and dynamics --- ISM:  H II regions --- ISM:
clouds --- Masers --- ISM:  individual (AFGL 5142) --- stars:  formation}

\section{Introduction} AFGL~5142 is a high-mass star forming region at a
distance of 1.8 kpc \citep{Snelletal1988}.  There are two centers of high-mass
star formation in this region:  IRAS 05274+3345 exhibits bright near infrared (IR)
emission \citep{hunter1995} with a far IR luminosity of $3\times 10^3$
\lsun, but has little dense molecular gas
associated with it \citep{estalellaetal1993}.  About 30$''$ to the east of
IRAS 05274+3345 (referred as IRAS 05274+3345 East) lies a faint cm continuum 
source of about 1 mJy at 3.6cm \citep{torrellesetal1992}.  If the continuum
emission arises from an optically thin, homogeneous \hii\ region, the
flux density is equivalent to a zero-age main-sequence (ZAMS) star of spectral
type B2 or earlier \citep{hunter1995,torrellesetal1992}.  The source is also
associated with emission from dust, dense molecular gas in \nh3, CS, HCN,
HCO$^+$, CH$_3$OH, and CH$_3$CN
\citep{estalellaetal1993,hunter1995,hunter1999,cesaroni1999,slyshetal1997,
pes_min_boo2005}, and \h2O, OH and class II CH$_3$OH masers
\citep{hunter1995,hunter1999,goddietal2005,slyshetal1997,pes_min_boo2005}.
High resolution images reveal a cluster of \h2O masers distributed in an area of
5$''$.  At a resolution of 3$''$, a 3mm continuum core is found with its peak
coincident with the cm continuum source \citep{hunter1999}.  All the
information indicates that IRAS 05274+3345 is a more evolved region, while the
active star formation occurs in the dense core 30$''$ to the east.

\citet{hunter1995} reported two outflows in the CO 2-1 line:  An extended
remnant flow in the southeast-northwest orientation possibly driven by the
brightest near infrared source near IRAS05274+3345, and a compact active outflow
in the north-south orientation originating from the cm/dust continuum emission
IRAS05274+3345 East.
This north-south outflow component is also seen in the CO 3-2 line observed
with the CSO \citep{hunter1995}.  Follow up studies at high angular
resolution from OVRO revealed a well collimated SiO jet and an HCO$^+$ outflow
in the north-south orientation \citep{hunter1999}.  In addition, there
appears to be a uni-polar outflow in the HCO$^+$ emission at a 35$^\circ$
position angle.  Both outflows appear to originate from the 3mm continuum peak
which is unresolved at $3''$ resolution.  Copious near infrared \h2 emission
knots are associated with the outflow lobes
\citep{hunter1999,chenetal2005}.  The high resolution \nh3 studies with
the VLA reveal a compact disk-like structure of $1''.2$ in size with large
line width of 6.4 \kms-1, and a velocity gradient over a scale
of 2000 AU \citep{zhang2002}.  This \nh3 structure coincides with
the peak of the cm and mm emission, and was interpreted as a rotating
disk surrounding a massive star.

However, with a spatial resolution of 3$''$, the mm observations 
from OVRO  may not resolve close binaries in a cluster environment.
In order to further investigate the outflow, and the structure and kinematics in
the core, we observed the IRAS 05274+3345 East region with the Submillimeter 
Array in CO and its
isotopologues, dense gas tracers and 1.3mm continuum emission at a
resolution of about 1$''$.  In addition, we also obtained the 3.6 cm continuum
image from the Very Large Array. The continuum images at 1.3mm and
3.6cm  indeed reveal
multiple emission peaks.  The CO and SO emission reveal at least three molecular
outflows.  In Section 2, we present the observational setup.  In Section 3, we
describe the main results revealed in the study.  And in Section 4, we discuss the
findings in the context of cluster formation.  We conclude in Section 5 with
the main findings of the investigation.

\section{Observations and Data Reduction}

\subsection{SMA}

Observations of AFGL5142 were carried out with the Submillimeter Array
\footnote{The Submillimeter Array is a joint project between the Smithsonian
Astrophysical Observatory and the Academia Sinica Institute of Astronomy and
Astrophysics, and is funded by the Smithsonian Institution and the Academia
Sinica.}  \citep{ho_moran_lo2004} on 2004 January 11 with six antennas in the
compact configuration, and on 2004 February 17 with 8 antennas in the extended
configuration.  The projected baselines of the two array configurations range
from 15m to 210m (11 to 160 $k\lambda$).  The double sideband receivers 
cover 2 GHz band width
at IF frequencies of $4 - 6$ GHz.  The receivers were tuned to the
LO frequency of 225 GHz to capture $^{13}$CO 2-1, C$^{18}$O 2-1, and CH$_3$CN
12-11 in the lower side band, and $^{12}$CO 2-1 in the upper side band.
The digital correlator was configured to provide a uniform channel spacing
of 0.8125 MHz ($\sim$ 1 \kms-1) across the entire band.  A detailed
description of the SMA is given in \citet{ho_moran_lo2004}.

The primary beam of the SMA at this frequency is about $57''$.  The phase
center for the observations is $\alpha (J2000) = 05^h30^m48^s.02$, $\delta (J2000)
= 33^\circ 47'54''.47$.  We used quasars 0359+509 and 0555+398 to calibrate time
dependent gains, and Jupiter to remove the gain variations across the
passband.  The flux scale was referenced to the Jovian moon Callisto.  The visibility
data were calibrated with the IDL superset MIR 
package\footnote{http://www.cfa.harvard.edu/$\sim$cqi/mircook.html}  
developed for the Owens Valley Interferometer.  The absolute flux level 
is accurate to about 15\%.  After the
calibration in MIR, the visibility data were exported to the MIRIAD format for
further processing and imaging.  The continuum is constructed from the line
free channels in the visibility domain, and is further self-calibrated 
using the clean components
of the image as input models. The gain solution from the self calibration
is applied to the spectral line data.  The rms in the naturally weighted maps is
4.4 mJy/beam in the continuum, and 90 mJy~beam$^{-1}$ per 1.2 \kms-1 channel
in the line data.  The synthesized beams of the images vary from $1''$ to
3$''$ (1700AU to 5100 AU) depending on the weighting of the visibilities in 
the Fourier transformation. The uncertainty in the absolute position
is $\lsim 0''.2$, derived from comparing quasar positions from two 
sidebands.


\subsection{VLA} 

Observations of AFGL5142 with the Very Large
Array \footnote{The National Radio Astronomy Observatory is operated by
Associated Universities, Inc., under cooperative agreement with the National
Science Foundation.} were carried out  on 2003 June 15 in the A configuration.  The correlator
was configured in the continuum mode at 3.6cm with a total bandwidth of 200
MHz.  The visibility data were made available to public through the
VLA data archive.  The
pointing center is the same as that of the SMA.  The primary beam of the VLA
is about $6'$ at this frequency.  3C147 was used as the flux calibrator, and
0443+346 was used as the gain calibrator.  The visibility data were calibrated
using the AIPS package.  The rms in the naturally weighted image is 0.09
mJy~beam$^{-1}$, with a synthesized beam of $0''.32 \times 0''.26$ 
(or 510 $\times$ 440 AU) at a
position angle of $58^\circ.6$. The uncertainty in the absolute position
is better than $0''.03$, one-tenth of the synthesized beam.

\subsection{Supplemental Short Spacing Information to the $^{12}$CO Emission}

The shortest projected baseline (11$k\lambda$) in the SMA observations corresponds to a spatial
scale of $1.2 {\lambda \over b}$, or 20$''$.  Here b is the length of the
projected baseline.  Spatial structures more extended than 20$''$ were not sampled in
the SMA observations.  This filtering effect can affect the appearance of
images of easily
excited molecular lines such as CO.  To recover the missing short
spacing information in the $^{12}$CO 2-1 line, we add the single dish data
obtained from the CSO \citep{hunter1995} to the SMA data, following a
procedure outlined in \citet{zhang1995}.  The CSO spectra have a velocity
range from -35 to 30 \kms-1, and do not cover the entire high velocity wings
detected in the SMA data.  However, the highest velocity CO emission appears to be
spatially compact, thus does not suffer strongly the missing flux
problem.

The combined image of the $^{12}$CO 2-1 line, when convolved to the 29$''$
beam of the CSO, recovers about 90\% of the flux from the single dish
telescope.  The $^{13}$CO 2-1 emission from the SMA also appears to miss extended
emission around the cloud systemic velocity.  For that reason, we will avoid
interpreting the $^{13}$CO image around the cloud systemic velocity.

\section{Results}

\subsection{Continuum Emission}

Figure 1 presents the continuum images at 225 GHz (1.3 mm) and 8.4 GHz (3.6 cm)
overlaid with
the \h2O maser emission from \citet{hunter1999}.  
The VLA image reveals three continuum peaks.  The two peaks, CM-1A and CM-1B
in the north are separated by $0.2''$.
They have a peak intensity of 0.65 mJy/beam and 0.64 mJy/beam,
respectively, well
detected above the $1 \sigma$ rms of 0.09 mJy/beam.  These two peaks were not
resolved in previous observations with the VLA.  The integrated flux of the
two sources are 1.4 mJy, consistent with observations at lower resolution
\citep{hunter1999}.  There appears to be an additional cm continuum source
2$''$ to the south of CM-1A/1B.  This source, labeled as CM-2, has a flux intensity of 0.35
mJy/beam.  CM-2 was not reported in the previous
observations possibly due to confusion from the extended emission.

At a resolution of $1''.3 \times 0''.8$, the 225GHz
continuum image reveals two dominant peaks separated by about $1''$.  The
northern peak, MM-1, coincides with two 8.4 GHz continuum peaks CM-1A/1B.
The southern peak, MM-2, does not have any counterpart at 8.4 GHz at a
$1 \sigma$ sensitivity of 0.09 mJy/beam.  

In addition to MM-1 and MM-2, there exist other mm continuum peaks. 
The extension toward the southwest from MM-2,
labeled as MM-3, is close to the the cm emission
feature CM-2. The \h2O maser feature $3''$ to the east of MM-1
coincides with a dust peak MM-4. Furthermore, a dust emission peak
lies about $2''$ west of MM-3. We refer to this peak as MM-5.
The flux density per beam at the continuum peaks are
listed in Table 1. 
The expected contribution from the ionized gas 
at mm and submm wavelengths is $\lsim 1$ mJy \citep{hunter1999}.
Thus, dust emission dominates the 1.3mm continuum flux of all
the sources, and likely does so to wavelengths as long as $\sim$ 3 mm.  
Because of a lack of high resolution images at more than one
wavelength, the emissivity index of the dust emission at each peak
cannot be evaluated.  Instead, we convolve the continuum map at 225 GHz to the
same synthesized beam as the 88 GHz map in \citet{hunter1999} to estimate
an average emissivity index for the region.  We obtain a peak flux density of
0.62 Jy at 225 GHz.  Comparing with the flux density of 0.038 Jy measured in the same beam at
88 GHz, we derived an index $\alpha = 3$, or an emissivity index of
$\beta$ = 1.  This $\beta$ is smaller than the typical value of 2 for the
interstellar dust, but is consistent with the value in a similar 
object IRAS 20126+4104 \citep{cesaroni1999}.
Using the Hildebrand dust opacity law \citep{hildebrand1983}, 
we find the dust opacity per unit dust mass $\kappa_0 (225GHz) = 1.8 $ \CM2/g.
For a dust
temperature of 45 K based on the estimate in \citet{hunter1999}, 
and an integrated flux of 1.5 Jy,
we obtain a total mass of 50 \msun\ in the region.  
We note that this mass is
smaller than 145 \msun\ derived from the 3mm flux in \citet{hunter1999}.  The
difference is attributed mostly to the dust opacity:
\citet{hunter1999} adopted an opacity of
0.5 \CM2/g at 230 GHz, which is about 30\% of the opacity used in this paper,
and hence resulted in a higher mass estimate.  
With the Hildebrand dust opacity, the circumstellar mass 
within one synthesized beam ($\sim$ 1800AU) estimated from the 225
GHz peak flux is 3 \msun\ for MM-1 and MM-2, respectively.
The kinetic temperature toward MM-3, MM-4 and MM-5 is about 20 K based
on the \nh3 studies \citep{zhang2002}. At that temperature,  
we derive a mass of 2 \msun\ for MM-3, 0.9 \msun\ for MM-4, and
0.7 \msun\ for MM-5, respectively.
  
 As shown in Figure 1, MM-1 appears to coincide with CM-1A/1B, and MM-3 
appears to coincide with CM-2. The maximum offset among the cm/mm counterparts 
is $<0''.2$, within the uncertainty in the astrometry of the SMA image. 
Therefore, the spatial displacement between the mm and cm peaks
may not be physical. The centimeter emission likely arises either from
the \hii\ region due to ionization of massive stars
or from the ionized wind emission in outflows. In general, the thermal free-free 
emission from the \hii\ region tends to have a negative spectral
index $\alpha$ (defined as $F_\nu \propto \nu^\alpha$) of about -0.1
in the optically thin limit. On the other
hand,  the emission from the ionized wind tends to have a positive
spectral index of about 0.6 \citep{angladaetal1998}. Without
images of similar resolution at other cm wavelengths, we cannot access
the physical nature of the emission. If the emission arises from the
optically thin and homogeneous ionized gas in the \hii\ region, 
the flux density from the three cm peaks corresponds to ZAMS stars of 
spectral type B3 to B2. However, as discussed in Section 4.1, the
3.6cm emission is probably due to ionized wind emission from
outflows.

\subsection{Line Emission}

\subsubsection{Line Spectroscopy}

Figure 2 shows the SMA spectra in the lower side
band and the upper side band toward
the position of MM-1, MM-2, MM-3 and the red-shifted peak of the
outflow A (marked as the triangle in Figure 3).  The data
are subtracted by the continuum emission constructed from the line-free
channels in the visibility domain, and then imaged with 
robust weighting, resulting in a synthesized beam of
$1''.3 \times 0''.8$ at a position angle of $-60^\circ$. 
Thus, the spectra taken from  the three dust continuum 
peaks are spatially resolved from each other.
In order to increase S/N, the spectrum toward the outflow A is
made with the natural weighting and a spatial tapering
of 3$''$, resulting in a resolution of $4''.3 \times 3''.6$
at a position angle of $-80^\circ$. 

Similar to other massive
star forming regions \citep{beuther2004g,beuther2004a}, multiple molecular
lines are detected in the AFGL 5142 region.  We detect 22 spectral lines/components 
from $^{12}$CO, $^{13}$CO, C$^{18}$O, CH$_3$CN, CH$_3$OH, SO,  HNCO, OCS and H$_2^{13}$CO
molecules (see Table 2).  The energy levels of the lines range from 17 K for CO
to 579 K for CH$_3$OH with
critical densities from 10$^2$ \cm3 to 10$^7$ \cm3.  These lines sample a wide
range of physical and chemical conditions in the region.  The peak in the
$^{12}$CO 2-1 line appears to be weaker than that of the $^{13}$CO. This is 
likely due to high optical
depths near the cloud LSR velocity and extended emission missed by the
interferometer.

As shown in Figure 2, the strength of the molecular lines varies significantly
in this region.  $^{12}$CO and SO are the only molecular lines
detected toward all four positions. MM-1 and MM-2 appear to emit
in nearly all the molecular lines identified in the region, including
hot core molecules CH$_3$CN and  CH$_3$OH. MM-3, a faint cm source
with a mass similar to that of MM-1 and MM-2, is not detected
in any of these hot core molecules. This suggests that MM-3 core has a lower
gas temperature than in the MM-2 and MM-1 cores. The presence
of SO toward the outflow indicates that
the SO abundance is enhanced greatly in the outflow as compared to the core.

Table 2 lists the flux and linewidth of the molecular lines detected toward
MM-1 and MM-2.

\subsubsection{Line Emission from the Core}

Figure 3 shows the images of the velocity integrated flux from
$^{13}$CO J=2-1, C$^{18}$O J=2-1, SO $6_5 - 5_4$, CH$_3$OH $8_{-1,8} -7_{0,7}$ E
at 229.759 GHz, OCS J=19-18, and HNCO $10_{0,10} - 9_{0,9}$ at 219.798 GHz.
Emission from all the molecules presented here is seen toward the core
surrounding MM-1 and MM2.  The $^{13}$CO emission  around the
cloud velocity appears to be
extended and likely suffers from missing large scale component. 
Thus, we will not discuss the emission in details.
The SO emission reveals an extended component corresponding to
molecular outflows and we will interpret the data in Section 3.2.3.
The emission from OCS and HNCO  is
associated with dense cores surrounding MM-1 and MM2 only, and shows the 
most compact structure. The CH$_3$OH emission likely arises from
both dense core and outflow.

The CH$_3$CN J=12-11 line also displays compact emission  
similar to that of OCS and HNCO. 
Figure 4 presents the integrated CH$_3$CN emission and the spectra toward
MM-1 and MM-2. At a resolution of $1''.3 \times 0''.8$, the two mm continuum 
sources are resolved from each other 
in the CH$_3$CN emission. As shown in the CH$_3$CN spectra,
the emission from MM-2 is  much stronger than that from MM-1,
thus dominates the integrated flux in Figure 4.
The hyperfine components of the K=0 to 6 of CH$_3$CN with 
energy levels from 69 to 326 K are detected.   The emission from the higher K 
components of CH$_3$CN toward MM-1
is much weaker  than that toward MM-2. This indicates
that MM-1 has a lower temperature than MM-2. Despite the
difference in statistical weights by a factor of 2, the K=2 and 3 components
from both sources have similar brightness temperatures,
indicating a moderate optical depth in the lines. 
Assuming LTE, we fit the CH$_3$CN 
line profiles using a radiative transfer model
in a uniform, isothermal spherical core \citep{chenetal2006}. The emergent
spectrum depends on the gas temperature, density, central velocity, 
linewidth and the size of the emitting region.
By varying these parameters, we obtain the best fit
by minimizing the $\chi^2$ between the observed and model spectra.
This approach takes into account
the optical depth effect, thus, produces a better fit
to the line profiles of all the K components than using the energy diagram which
assumes optically thin emission (e.g. \citep{zha_ho_oha1998}). 
Figure 4 shows the CH$_3$CN J=12-11 lines reproduced from the model: The
match between the observed and model spectra is reasonable. 
The fitting yields \vlsr\ = $-1 \pm 0.2$ \kms-1, FWHM = $4.7 \pm 0.5$ \kms-1, 
$T_K$ = $90 \pm 20$ K, n$_{CH_3CN}$ = $1.3  \pm 0.7$ \cm3\ and the core radius
R = $440 \pm 80$ AU
for MM-1, and \vlsr\ = $-3.4 \pm 0.1$ \kms-1, FWHM = $4.0 \pm 0.2$ \kms-1, 
$T_K$ = $250 \pm 40$ K,  n$_{CH_3CN}$ = $6 \pm 4$ \cm3\ and R = $380 \pm 80$ AU
for MM-2. 
The central velocity of the CH$_3$CN cores toward MM-1 and MM-2 differs
by 2.4 \kms-1. This difference, also seen in the SO and
CH$_3$OH emission, confirms that both dust continuum sources
contribute to the observed CH$_3$CN emission.

The temperature and density derived from CH$_3$CN reveal that the physical
conditions in the MM-1 and MM-2 cores are different. MM-2 has much higher
temperature and column density than MM-1. As shown in
Table 2 and Figures 2 and 4, the higher K components of the CH$_3$CN line
toward MM-1 are weaker than toward MM-2.
In addition, emission from higher energy levels of CH$_3$OH and HNCO are much 
weaker or not detected toward MM-1. These measurements are consistent with a lower
temperature and a lower CH$_3$CN density toward MM-1. From the dust emission,
MM-1 and MM-2 have a similar peak flux. At the assumed dust temperature
of 45 K, the mass in both cores (3 \msun) yields 
a mean \h2 density of $1.5 \times 10^8$ \cm3 averaged over the volumn of the 
synthesized beam (1800AU). Thus, the fractional abundance of CH$_3$CN
relative to \h2\ is $1 \times 10^{-8}$ and $4 \times 10^{-8}$ 
toward MM-1 and MM-2, respectively.
Both values are two orders of magnitude higher than the abundance
in dark clouds \citep{HerLeu1990}, similar to that in the hot core \citep{wilner1994}. 
The abundance in the MM-2 is
higher than in the MM-1 core.  Based on the fitting of the CH$_3$CN spectra,
MM-2 has a gas temperature 2.5 times of that in the
MM-1 core. If the gas and dust reach an equilibrium temperature,
the mass and hence the average \h2 density in MM-2 
should be smaller than in MM-1. This would result in even higher CH$_3$CN abundance
toward MM-2. The higher CH$_3$CN abundance in MM-2 is likely
the result of higher temperature in the MM-2 core, which enhances the
production of CH$_3$CN.

In addition to CH$_3$CN,  emissions from SO,  OCS, HNCO and CO
isotopologues are also detected toward MM-1 and MM-2. The central
velocity of the emission from these molecular lines is centered 
around -1 and -3.4 \kms-1\ for MM-1 and
MM-2, similar to the that in the CH$_3$CN emission.
As shown in Section 3.2.3, the SO and CH$_3$OH emission can arise from outflows.
The agreement of the line
velocity indicates that the SO and CH$_3$OH emission toward
MM-1 and MM-2 arise from the dense core.

Unlike MM-1 and MM-2, far fewer molecular lines are detected toward MM-3.
In particular, CH$_3$CN, CH$_3$OH, OCS and HNCO are all absent (at a 
1$\sigma$ sensitivity of 0.09 Jy), despite the presence
of faint 8.4 GHz emission. The absence of these lines also indicates
that MM-3 is at a lower temperature than that of
the MM-1 and MM-2 cores.

\subsubsection{Molecular Outflows}

Figure 5 presents the channel maps of the $^{12}$CO 2-1 line.  Since the single
dish CO 2-1 data from the CSO do not cover the high velocity wing emission,
the combined SMA and CSO data have a limited velocity coverage from -35 to 30
\kms-1.  The CO emission at $\Delta V >$ 15 \kms-1\ is compact and subjects to little
missing flux.
Thus, we show in Figure 5a the high velocity emission obtained from the
SMA alone. Figure 5b presents the low velocity CO emission from the
combined SMA and CSO data.  From the CO emission, one can immediately
identify an outflow in the north-south direction.  The CO 2-1 emission 
from the CSO \citep{hunter1995} peaks
at -3.9 \kms-1.  At the red-shifted velocities of 10 to 48 \kms-1, there
exists high velocity gas toward the north of the mm continuum peaks at a
position angle of $5^\circ$.  The well-collimated CO emission is detected up
to 50 \kms-1\ from the cloud systematic velocity and extends up to $30''$
from the mm continuum sources.  The velocity of the red-shifted CO emission
increases with the distance from the mm
continuum sources.  The red-shifted lobe appears to be clumpy.  The
peak of the CO clumps is not aligned along the outflow axis and spans an angle of $\sim 5^\circ$
with respect to the dust continuum peaks (see channels from 12 
to 45 \kms-1\ in Figure 5a).
The blue-shifted side of the outflow extends about $5''$ to the south
from the mm continuum peaks, with the CO emission detected up to -13 \kms-1.  This
outflow, referred as outflow A, coincides spatially with the high velocity SiO
outflow reported in \citet{hunter1999}.  

Additional high velocity CO emission suggests the presence
of other outflows. At the blue-shifted
velocities of -14 to -26 \kms-1, the CO emission at a position
angle of $35^\circ$ identifies the outflow B.  The lobe to the southwest 
of the mm continuum sources
is mainly blue-shifted with CO emission detected up to -22 \kms-1\ from the
cloud velocity.  The lobe to the northeast exhibits both blue- and red-shifted
emission from -18 to 15 \kms-1.  Weak CO emission is detected about $30''$ southwest
of the mm continuum peaks.  The possible third outflow, outflow C, originates from
the mm core at a position angle of $-60^\circ$.  The lobe to the
northwest of the mm continuum peaks has a bifurcated structure at -8 \kms-1,
extending $20''$ to the northwest.  Higher velocity emission of $\pm$25 \kms-1\
is seen toward one side of the shell.  The lobe to the southeast has a lower
velocity with a jet like feature seen in the -8 \kms-1 channel.

Furthermore, there appear to be additional high
velocity features that may arise from other outflows. For example,
the CO emission around [$-15'', -15''$] from MM-2 seen
at velocity channels of -16.8 and -10.4 \kms-1\ does not align with any of the
three outflows identified and may trace a fourth outflow in the region.

Figure 6 presents the position velocity plots of the three outflows identified
above.  To show the full velocity extent of the outflows, we use the image
from the SMA data only.  The negative contours (dotted lines in Figure 6) near the
cloud velocity arise from the effect of missing short spacing flux.  Figure 6
reveals that the maximum velocity in the outflow A increases with the distance
from the driving source.  This `Hubble'-law relation has been seen in
molecular outflows toward both high- and low-mass stars
\citep{su_zha_lim2004,lad_fic1996,gue_gui1999}.  Outflows B and C show rather
complex velocity structures, with both the blue- and red-shifted high
velocity gas toward the same side of the lobes. This may
indicate additional outflows in the region.

Figure 7 shows the channel maps of the SO $6_5 - 5_4$ transition.  
SO is present in dark clouds \citep{swade1989}, and its abundance is
enhanced in outflows \citep{bachilleretal2001,vitietal2003}. The 
process involves the photo
evaporation of sulphur bearing molecules such as \h2S from dust grains
near shocks.
The gas phase \h2S then reacts with OH and O$_2$ to produce SO and SO$_2$.
Strong SO emission
is detected toward the CO outflows in AFGL5142, but at velocities
much closer to the cloud velocity: The SO emission is
detected within a velocity range of -15 to 8 \kms-1, while the CO emission
is seen between -26 and 50 \kms-1.  Despite the velocity
difference from the CO emission, SO emission appears in
all three molecular outflows.  Outflow A
is seen in both the red-shifted (-3.8 to 5.8 \kms-1) and blue-shifted
emission (-3.8 to -7.4 \kms-1).  The spatial extent of the blue-shifted and
the red-shifted lobes is similar to that of the CO outflow.  The outflow B is
clearly seen in SO as well:  The red-shifted emission in the northeast is
detected from -8.6 to 5.8 \kms-1, while the blue-shifted emission to the
southwest is from -3.8 to -12.2 \kms-1.  For outflow C, the SO emission is
mostly seen from 2.2 to -8.6 \kms-1\ for the lobe in the southeast, and 5.8 to
-2.6 \kms-1 for the lobe in the northwest.

In addition to SO, the CH$_3$OH emission is detected in the outflows as well.
Although CH$_3$OH abundance is low ($< 10^{-9}$) in dark clouds,
UV radiation from outflow shocks can release CH$_3$OH from dust grains, 
and enhance the CH$_3$OH abundance
in molecular outflows \citep{bachilleretal2001,vit_wil1999}.  Figure 8 presents
the channel maps of the CH$_3$OH transition $8_{-1,8} - 7_{1,6}$ E. The strongest
emission is detected along outflow B 
at velocity channels from -0.2 to -7.4 \kms-1.  In addition, an extension
to the northwest from MM-1, MM-2 and MM-3 appears to be associated
with outflow C. In general, the CH$_3$OH emission in the outflow is much weaker
than the SO emission and appears at lower velocities than CO and SO.
Furthermore, the CH$_3$OH emission
does not always coincide with the SO emission spatially.

From the CO emission, we estimate the mass and momentum in the outflows.  Many
studies of molecular outflows assumed optically thin CO emission when
estimating outflow parameters.  In some cases, an average value of 10
for the $^{12}$CO to $^{13}$CO ratio is assumed to
correct for the optical depth effect \citep{cho_eva_jaf1993}.  
However, the opacity is in general a function of velocity.
As shown in \citet{su_zha_lim2004}, the CO can be optically thick even 
at high velocity wings.  
The simultaneous observations of the $^{12}$CO and $^{13}$CO with the
SMA allow us to investigate the CO optical depth in the outflow. Although
$^{13}$CO suffers missing flux near the cloud velocity, the channel
images suggest that such an effect
is negligible for emission with $\Delta V > 4$ \kms-1. By examining
the emission ratios, we found that
the CO optical depth varies from 4 to 15 in the line wings 5 to 12
\kms-1 from the cloud velocity.  Making
appropriate corrections for the opacity, we find a total mass, momentum and
energy of 5.9 \msun, 38 \msun~\kms-1, and $3.4 \times 10^{45}$ ergs,
respectively, for the red-shifted CO emission from 1.6 to 50 \kms-1.
Similarly, we integrate the blue-shifted CO emission from -8 to -40 \kms-1,
and find a mass, momentum and energy in the outflow of 5.7 \msun, 32
\msun~\kms-1, and $2.5 \times 10^{45}$ ergs, respectively.  There are about
equal amounts of mass in the three outflows: 
2.5, 3 and 3 \msun\ in the outflows A, B and C, respectively.  The momentum
in the outflows are from 12 to 20 \msun~\kms-1.  The dynamic time scales of
the outflows, computed from $l \over V_{max}$ are about
$10^4$ yrs. Here $l$ and $V_{max}$ are the maximum length and
velocity of the outflow where
the CO emission is detectable at an rms of 0.09 Jy.
The outflow parameters for the three outflows are listed in Table 2.
Due to the existence of additional high velocity features,
the masses from the three outflows do not add up to the total
outflow mass for the entire region.

\section{Discussions}

\subsection{Nature of the Continuum Sources}

The presence of multiple continuum peaks and a group of \h2O masers indicates a 
deeply embedded cluster in formation.  Both MM-1
and MM-2 give rise to CH$_3$CN emission with rotational temperatures of 90
and 250K, respectively. Although nearby low-mass protostars may excite hot-core type 
molecular emission \citep{ceccarellietal2000,kuanetal2004,chandleretal2005},
the emission may not be detectable at larger distances.  
The protostars that produce heating toward MM-1 and MM-2 and the 
CH$_3$CN emission are likely to be massive. 
\citet{sco_kwa1976} derived a relation between
the dust temperature and the luminosity of the star:  $$ T_D = 65 ({0.1pc\over
r})^{2/(4+\beta)} ({L_{star}\over 10^5 L_\odot}) ^{1/(1+\beta)} ({0.1\over
f})^{1/(4+ \beta)} K.$$ Here $\beta$ is the power law index of the dust
emissivity at the far infrared wavelengths, $f$ = 0.08 cm$^2$~g$^{-1}$ is the
value of the dust emissivity at 50$\mu$m, and $r$ the distance from the star.
The dust temperature derived from the fitting to the SED using a grey body
assumption gives a dust temperature of 45 K \citep{hunter1999}.  This
value corresponds to the average temperature over $>20''$ scales.  At scales of 1$''$,
dust temperatures probably reach as high as 250K, the gas temperature derived from the
CH$_3$CN emission, if the gas and dust reach thermal equilibrium in the high
density environment.  With $T_D = 45 - 200$ K, $r = 900 $ AU the size
of dust emission, and $\beta = 1$,
we estimate the luminosity of the star of $4 \times 10^3$ to 
$7 \times 10^4$ \lsun.  Thus, both protostars
are likely to be massive based on their luminosity. MM-2, with higher
temperature and CH$_3$CN abundance may be more evolved than the source
embedded in the MM-1 core.

The central velocity of the molecular gas in the two hot cores is
offset by 2.4 \kms-1. If MM-1 and MM-2 are in a binary,
the binding mass of the system is 10/$sin(i)$ \msun, with
$i$ being the inclination angle of the orbit.  The mass in both cores adds up
to 6 \msun.  If the two protostars are early B type as suggested by the
presence of hot cores, the combined stellar mass can be up to 20 \msun.  
Therefore, the total
mass in the stars and circumstellar material is large enough to host the
binary.  If so, the projected separation of the two stars
is 1700 AU with an  orbital period of about $10^4$ yrs.  

 Dynamical interactions between members in a proto binary system 
can cause  outflow precession. The CO peaks toward
the red-shifted lobe of Outflow A
exhibit misalignment along the outflow axis, consistent with a 
precessing jet.  Jet precession has been reported
in protostellar outflows associated with
both low- and high-mass young stars \citep{zhang2000,shepherd2000}.
It appears that the precession of $5^\circ$
in the outflow A is much smaller than the $20^\circ - 45^\circ$  reported
in some of the other protostellar systems 
such as IRAS 20126+4104 \citep{shepherd2000,lebron2006}.

MM-3 appears to be associated with the 8.4 GHz emission, but does not 
have detectable
organic molecules indicative of a hot core. If the 8.4 GHz emission arises
from the \uchii\ region, the central star that photoionizes the circumstellar 
material should also give rise to high temperatures in the molecular
gas. The absence of a hot core toward MM-3 indicates a relatively
lower temperature in the molecular gas. This fact leads us to
suggest that the faint 8.4 GHz emission
likely arises from the ionized wind emission driven by an 
intermediate to high mass protostar in the MM-3 core.
Although the ionized wind emission
has been detected toward nearby low-mass young stars,
the typical flux is on the order of 0.1 to a few mJy \citep{angladaetal1998}.
Such flux levels, after being scaled down by a factor of 10 to 100 to account
for the distance,
would not be detectable at  the distance of AFGL5142 at an 
rms of 0.09 mJy/beam. 
Thus, the ionized wind emission toward MM-3 is likely to be powered by a
more massive young star. In a model in which the ionized emission arises
from the plane parallel shock when the neutral wind plows into
the high density material, \citet{curieletal1989} derived a relation between the
momentum rate in the outflow and radio flux:
$$({\dot{P} \over M_\odot yr^{-1} km~s^{-1}}) = {10^{-3.5} \over \eta} ({S_\nu
\over mJy}) ({d \over kpc})^2. $$
Here $\dot{P}$ is the momentum rate in the outflow, $\eta = 0.1$ is the efficiency
factor corresponding to the fraction of the stellar wind that is shocked and
produces the radio emission. For a flux $S_\nu$ of 0.35 mJy, we obtain an
outflow momentum rate  $\dot{P} = 4 \times 10^{-3} $ \msun yr$^{-1}$ \kms-1.
This is in rough agreement with the outflow momentum rate in the
CO outflow. In order to test our hypothesis on the ionized wind nature of
CM-2, more sensitive cm continuum observations are needed to measure its
spectral index.

Additional evidence that MM-3 is powered by an embedded massive star comes from
the amount of circumstellar mass present.
Although the total mass in MM-3 appears to be smaller
as compared to MM-1 and MM-2, the mass estimates from the dust emission
depend (1/T) on the dust temperature assumed. We assumed a dust temperature
of 45 K for MM-1 and MM-2.  If we use the gas temperature of 250 K for MM-2, the
mass in MM-2 would be 0.7 \msun, comparable to the mass in MM-3. Thus,
many lines of evidence suggest that MM-3 is an intermediate to massive 
star at an earlier
evolutionary stage than MM-1 and MM-2. At such a stage, the radiation of
the central star may not yet have produced a large enough core with high
temperature that renders detection in hot core molecules.

In addition to the strong dust peaks MM-1, MM-2 and MM-3, MM-4 coincides with
the \h2O maser feature $3''$ to the east of the strong mm peaks.
\h2O masers can be
excited in dense envelopes ($\sim 10^8$ \cm3) surrounding 
protostars in the interaction with molecular 
outflows \citep{Eli_Hol_Mck1989,felli1992}.
Thus, the \h2O maser emission may indicate the
presence of a fourth young star.

In contrast to the spatial agreement of a water maser with MM-4, a
pair of \h2O masers lie near symmetrically $2''$ to the southeast 
and northwest of MM-5. There appears to be a high velocity CO feature
(see the channel maps at velocities from 4 to 8.8 \kms-1 in Figure 5b)
aligned with this source. This CO feature can be part of outflow C associated
with MM1. On the other hand, it could also be a separate outflow
associated with MM-5.

\subsection{Kinematics and Driving Source of Outflows}

Of the three molecular outflows identified in the CO and
SO emission,  outflow A coincides with a jet-like SiO outflow reported by
\citet{hunter1999}.  Figure 9 presents an overlay of
the position-velocity diagram
of the CO and the SiO emission along the axis of the outflow A.  
For the SiO 2-1 line, strong emission
appears around $\pm 6$ \kms-1\ from the cloud systemic velocity.  Higher
velocity emission is detected up to $\pm 40$ \kms-1\ from the cloud velocity
near the position of the driving source, $i.e.$, the position offset around $0''$.
The maximum velocities of the SiO emission along the outflow axis
increases with the distance from
the driving source, similar to that of the CO outflow.

Outflowing gas moving faster than the sound speed
in a cloud (a few \kms-1) produces shocks when interacting with the core 
material and
enhances the gas temperature.  As reported in
\citet{zhang2002}, gas heating due to the outflow A is seen in AFGL5142.
We re-examined the ratio of the \nh3 (J,K)=(3,3) and (4,4)
to the (1,1) line presented in \citet{zhang2002}.  
Figure 10 shows the \nh3 ratio maps with all three outflows
marked. The \nh3 (J,K) = (1,1), (3,3) and (4,4)
emission  are from energy levels of 23, 125 and 230 K, respectively.
Since upper (J,K) transitions arise from  higher energy levels,
the higher (3,3)/(1,1) and (4,4)/(1,1) ratios reveal warmer
molecular gas.
It is clear that in addition to the temperature enhancement along the
outflow A, there appears to be enhanced temperatures along outflows B and C.

All three outflows appear to originate from the central core of about 3$''$
area.  Therefore, it is difficult to identify the driving source.  Using the
data in this paper, and the \h2O maser distribution and proper motion, 
we attempt to assign driving sources for the three outflows with a caveat that the
assignment may not be unique.  There are
three cm continuum peaks within the 3$''$ region.  CM-1A and CM-1B are
separated by
less than $0''.2$ and coincide with the 225 GHz continuum peak MM-1.
The \h2O maser studies with the VLBA by \citet{goddietal2005} show a 
cluster of masers associated with the CM-1A/1B and MM-1.  The
masers are concentrated to the northwest and southeast of the continuum
source and exhibit proper motion along the axis of PA=-40$^\circ$.  This
orientation is close to the axis of the outflow C.  Therefore, it is likely
that MM-1 is the driving source of the outflow C.  The two cm continuum
peaks toward MM-1 can be  a close binary with a projected separation of 
400 AU. On the other hand, CM-1A and CM-1B are
aligned in a similar direction as the \h2O maser proper motion.
It is more likely that the two centimeter
peaks trace an ionized jet  in the outflow.

For the outflow B, both the CO and SO lobes (see the channel at -12.8 \kms-1 in CO)
appear to put the MM-3 close to the geometric center.  The \h2O maser
study by \citet{goddietal2005} also detected a pair of the maser spots at a
similar position
angle.  However, the geometric center of the \h2O masers appears to be
north of MM-3, and closer to MM-2.  This pair of masers was not detected in
other VLA observations \citep{hunter1999},  thus, their proper motion 
may be questionable.  We propose that
MM-3 is the driving source of outflow B.

\h2O masers are also detected in the vicinity ($<0''.5$) of MM-2.  This source,
not detected at 8.4 GHz with the VLA at an rms of 0.09 mJy/beam,
is associated with a hot core with strong CH$_3$CN emission.  Being the only
dominant continuum source remaining, MM-2 probably drives the north-south
outflow A.

\subsection{Chemical Variations}

The AFGL5142 region reveals interesting variations in molecular line emission.
SO emission is seen toward cores MM-1, MM-2 and MM-3, while CH$_3$OH,
CH$_3$CN, OCS and HNCO emission are detected only toward two hot cores
MM-1 and MM-2. This difference indicates that SO is abundant during
a long period of core evolution from an
early and relatively cold stage through a more 
evolved and warm/hot stage. On the other hand, CH$_3$OH and
CH$_3$CN are abundant only when massive cores are more evolved such that
heating from the central protostar evaporates the molecules from the dust.

SO and CH$_3$OH emission are also detected toward molecular outflows away from
the dense core. This is
expected as their abundances are enhanced in outflow shocks where heating
and UV radiation from shocks release CH$_3$OH and sulphur bearing molecules 
from dust, and enhance the formation of SO in the gas phase. The time scale
of the abundance enhancement is $10^4$ yrs \citep{vit_wil1999}, 
consistent with the dynamical time
scale of the outflow.

\subsection{Cluster Star Formation}

The multiple cm/mm continuum sources identified in the AFGL 5142 region
indicate a dense cluster in formation. The infrared imaging in J, H and K
by \citet{hunter1995} reveal 30 point sources with IR excess over 
an area of $3'$. However, nearly
all the IR sources lie outside of the dense core. The density of the IR sources
within the 0.1 pc radius from the mm continuum sources is about $10^3$ stars 
pc$^{-3}$. This is much smaller than
the average stellar density of $\lsim 10^5$ pc$^{-3}$ within the 
central 0.2 pc in Orion Nebula, with 3 O stars within
0.1 pc \citep{hillenbrand1997,hil_har1998}. W3 IRS5, a cluster with 
a total luminosity  of $2 \times 10^5$ \lsun,
has 5 near IR sources within 5600 AU, equivalent to $6 \times 10^5$ stars pc$^{-3}$.
Toward AFGL 5142, 3 protostars are
revealed in the mm observations with a projected separation of $2''.3$,
or 4100 AU, which is equivalent to a stellar density of $2 \times 10^5$ 
stars pc$^{-3}$. Because the average is performed over a very small
region, this high stellar density is not necessary representative of the
entire region. Nevertheless, the AFGL 5142 region appears to form
a dense cluster. The luminosity for AFGL5142
is 1-2 orders of magnitude smaller than that of the Orion Nebula and
W5 IRS5, and appears to harbor only early B stars. Unlike the
Orion Nebula and W3 IRS 5, however, all three protostellar
objects in AFGL 5142 appear to be actively accreting  because of
the presence of outflow activity. 
It is possible that they may become O stars when the star formation
is complete. 

\citet{zhang2002} reported a velocity gradient and
large \nh3 linewidth of 6.4 \kms-1\ toward the
peak of the 3mm continuum detected by \citet{hunter1999}. The kinematics
in \nh3 appears to be consistent with a rotating disk in the region. Higher resolution
observations with the SMA resolved the 3mm dust peak into two sources
separated by 1$''$, which excite two hot cores detected in the 
CH$_3$CN emission.
The two hot cores have systemic velocities of -1 and -3.4 \kms-1,which 
creates the velocity gradient seen in the \nh3 emission at lower resolution. Re-examining
the data in \citet{zhang2002}, we found that at $\sim 1''$ resolution 
the \nh3 (3,3) emission shows an elongation in the north-south direction.
There appears to be two velocity components in the \nh3 (3,3) channel maps
(Figure 2c in Zhang et al. 2002), consistent with the CH$_3$CN data. 
Refitting the $1''$ resolution
\nh3 (3,3) line yields a FWHM of $ \gsim 6$ \kms-1 for MM-1 and MM-2.
The linewidth is still significantly larger than the typical \nh3 linewidth
of 2 \kms-1 in the extended emission. The increase of linewidth
in \nh3 and CH$_3$CN is consistent with the enhanced
infall/rotation in the circumstellar material surrounding the star. However,
due to the dominance of velocity shift between the MM-1 and MM-2 cores,
no rotational motion toward the individual cores can be discerned over 
a scale of $> 1''$.

The study of the AFGL 5142 region highlights the complex kinematics in
cluster forming regions
where multiple massive stars compete for gravitational influence. 
In such an environment, extended rotating disks are likely truncated 
by the tidal interaction. The relative
motion of dense cores surrounding massive stars can dominate the kinematics,
and even create false signal of rotation. High angular resolution observations
in continuum and spectral lines are the key in understanding the
kinematics in these regions.

Despite the complexity in the core, the molecular outflows seen in CO and SO 
at larger scales provide insight
in the formation process of stars in the cluster environment.
By physical connection, the well collimated outflows indirectly
suggest the presence of circumstellar accretion disks toward the three 
protostars in this cluster environment.  One issue under
debate in the past few years has been the formation mechanism of massive stars.
 In contrast to low-mass stars that stellar radiation exerts
little dynamical influence  on the infalling material, 
strong radiation from massive stars becomes a barrier
to infall and may even prevent the formation
of stars beyond 8 \msun\ \citep{larson1969,larson1971,
kahn1974,yorke1977,wolfire1987}. Several
solutions have been proposed to alleviate this radiation problem.
The most effective solution is a flattened accretion 
disk \citep{nakano1989,jijina1996,yorke2002}
which shields most infalling material from the stellar radiation.
An additional benefit of a flattened disk is that photons can escape
through the lower density region along the polar axis of the 
disk \citep{yorke2002,krumholz2005a}.
All the above mentioned solutions build
upon the notion that massive stars form through gravitational
collapse and disk accretion process \citep{mckee2002,keto2003}.
\citet{bon_bat_zin1998} and \citet{bonnell2002} put forward an alternative process in which
massive stars form through multiple mergers/collisions of low-mass
protostars.  This coalescence model appears to circumvent the radiation problem,
but requires exceedingly high stellar density ($\gsim 10^6$ stars~pc$^{-3}$)
in order for the process to be effective. In addition, the model
still faces the radiation problem if the mechanical energy of the merging
stars is released through radiation \citep{bally2005}.  These two schools
of models can be tested by observations \citep{zhang2005b,cesaroni2006}. 
Unlike the infall/accretion model,
the coalescence scenario does not produce well
collimated outflows from massive young stars \citep{bally2005}.  
Our observations indicate that the coalescence model
does not appear to be at work in this high-mass star formation region, since
collimated outflows point to an alternative process:  $i. e.$ disk
mediated accretion.
At the same time, the coalescence model requires high stellar density ($\gsim
10^6$~pc$^{-3}$) for merger to be effective. 
This stellar density requirement is not likely met in this region.

The coalescence scenario is a variation of the competitive accretion
model \citep{bonnell2002,bonnell2004} in which a cluster of low-mass protostars with a common
gravitational potential accretes the distributed gas from a reservoir of material
in molecular cloud. Protostars located near
the center of the potential accrete at a 
higher rate because of a stronger gravitational pull, and
experience faster mass growth. This model explains the stellar
initial mass function observed \citep{bonnell2002,bonnell2004}. The main difference
between the turbulent accretion and competitive accretion is that 
in the former model protostars accrete the 
gravitationally bound gas, whereas in the
competitive accretion model massive protostars accrete mostly from the 
gravitationally unbound gas. As a consequence, the competitive
accretion model predicts that the final mass of massive stars does not correlate
with the core mass \citep{bonnell2004}. In addition,
the accretion rate in the turbulent accretion model goes with the turbulent 
velocity $V_t$ as $V_t^3/G$, while in the competitive
accretion model the accretion rate goes as $4 \pi \rho {(GM)^2 \over V_t^3}$ 
for Bondi-Hoyle accretion. Here G is the gravitational constant,
$\rho$ the gas density, and $M$ the mass of the accreting star.
We estimate the accretion rate using the parameters derived here.
For $V_t = 2.5$ \kms-1, $M = 5$ \msun, and $\rho = 10^5$ \cm3 (the mean density
in the core), we
find an accretion rate of $4 \times 10^{-3}$ \msun$~yr^{-1}$ for
turbulent accretion, and $2 \times 10^{-6}$ \msun$~yr^{-1}$
for Bondi-Hoyle accretion. The rate of $4 \times 10^{-3}$ \msun$~yr^{-1}$
is sufficiently high to account for the mass loss
in the outflow, whereas the rate of $2 \times 10^{-6}$ \msun$~yr^{-1}$
appears to be too low to account for the mass loss rate in the outflow.
On the other hand, the accretion rate is sensitive to turbulent velocity.
The measured line width reported in Table 2 is probably affected by 
systematic motion (infall/rotation and outflow) in the core. If assuming
a smaller value of $V_t = 1.5$ \kms-1\ based on the \nh3 line width in the 
extended core \citep{zhang2002}, we find an accretion rate 
of $8 \times 10^{-4}$ \msun$~yr^{-1}$ for turbulent accretion, 
and $1 \times 10^{-5}$ \msun$~yr^{-1}$ for Bondi-Hoyle accretion.
The latter value is still somewhat low, but closer to the accretion rate
required for producing the observed outflows \citep{zhang2005a}. 
Although the test of the two models based on the
AFGL 5142 data appears to be inclusive, it is hopeful that future
observations of massive cores at earlier evolutionary stage will provide 
measurements of
relevant physical parameters, especially turbulent velocity, and thus 
provide more definitive test of the two models.

\section{Conclusion}

High resolution ($1''$) observations with the SMA identified a high-mass protocluster
in AFGL5142. The 1.3mm image reveals 5 continuum peaks with large variations
in molecular line emission. Two mm cores, MM-1 and MM-2,  are
associated with the CH$_3$CN emission with rotational temperatures of
$\sim$ 90 and 250 K, and   abundance of $1 \times 10^{-8}$ and
$4 \times 10^{-8}$, respectively. The higher temperature and inferred high
luminosity indicate that they like form massive protostars. The remaining
three cores do not have detectable CH$_3$CN and CH$_3$COH emission, and likely
harbors less massive (or less evolved) protostars,

The CO and SO emission reveal at least three possible molecular outflows.  
The terminal velocity
of the CO emission reaches up to 50 \kms-1.  The outflows have masses of 3
\msun, and momentum of 12-20 \msun~\kms-1, typical of those associated with
high-mass stars.   Both CO
and SiO emission in the dominant outflow A have terminal velocities 
increase with the distance 
from the star.  The presence of
multiple collimated outflows in this protocluster provide indirect evidence 
that accretion is the
dominant process for the formation of the massive protostars
in this cluster.

\acknowledgements We thank P. Caselli for insightful discussions. H.B. acknowledges 
financial support by the Emmy-Noether-Programm of
the Deutsche Forschungsgemeinschaft (DFG, grant BE2578).

\clearpage

\begin{table}
\caption{Parameters of Continuum Sources$^{a}$} 
\begin{center}
\begin{tabular}{rllcc} \hline \hline
Name & R.A.$^{a}$(J2000)   & Dec.$^{a}$(J2000)  & $S_{peak}$ & Mass$^{b}$ \\ \hline 
     &  ($^h$ $^m$ $^s$)     &  ($^\circ$ $'$ $''$)    & mJy/beam   &  \msun\ \\ \hline 
CM-1A & 05 30 48.014 & 33 47 54.66 & 0.65 & - \\ 
CM-1B & 05 30 48.025 & 33 47 54.41 & 0.64 & - \\ 
CM-2 & 05 30 47.999 & 33 47 52.57 & 0.35 & - \\ 
MM-1 & 05 30 48.037 & 33 47 54.55 & 240 & 3 \\ 
MM-2 & 05 30 48.034 & 33 47 53.72 & 230 & 3 \\  
MM-3 & 05 30 47.979 & 33 47 52.42 & 84  & 2 \\  
MM-4 & 05 30 48.193 & 33 47 54.62 & 35 & 0.9 \\ 
MM-5 & 05 30 47.849 & 33 47 52.20 & 27 & 0.8 \\ \hline 
\end{tabular}
\tablenotetext{a}{There appears to be an offset of $0''.2$ between the 225GHz data, and the
8.4 GHz data and \nh3 data from the VLA.  This offset is within the calibration
error in the 225GHz dataset. Minor mm peaks less than 6$\sigma$ are 
not reported here.} 
\tablenotetext{b}{We assume dust temperature of 45 K for MM-1 and 
MM-2 based on the estimate in Hunter et al. (1999), and 
20 K for MM-3, MM-4 and MM-5 based on the \nh3 
observations (Zhang et al. 2002)}. 
\end{center} 
\end{table}

\clearpage

\begin{table}
\caption{Lines Detected in the Hot Core$^{a}$} 
\begin{center}
\begin{tabular}{lrr|rr|rr} \hline \hline
          &      &                           & MM-1 &  & MM-2 &  \\ \hline 
Frequency & Line & E$_{upper}$               & S & FWHM & S & FWHM   \\ \hline 
GHz       &      & K                         & K & \kms-1\ & K & \kms-1\ \\ \hline 
219.560 & C$^{18}$O 2-1                      & 16 & 9.4 & 2.9 & 11.7 & 4.4\\ 
219.733 & HNCO 10$_{2,9}$ - 9$_{2,8}$        & 230 & 4.5 &  - & 8.0 & 4.8 \\ 
219.737 & HNCO 10$_{2,8}$ - 9$_{2,7}$        & 230 & 4.5 & - & 8.0 & 4.8 \\ 
219.798 & HNCO 10$_{0,10}$ - 9$_{0,9}$       & 59 & 4.8 & 5.8 & 12.4 & 4.9 \\ 
219.909 & H$_2^{13}$CO 3$_{1,2}$ - 2$_{1,1}$ & 23 & 3.4 & 3.7  & 5.4 & 3.7 \\ 
219.949 & SO 6$_{5}$ - 5$_{4}$               & 35 & 20.2 & 5.6 & 30.0 & 5.8 \\ 
220.079 & CH$_{3}$OH 8$_{0,8}$ - 7$_{1,6}$   & 96 & 14.2  & 4.9 & 25.3 & 3.6 \\ 
220.399 & $^{13}$CO 2-1                      & 16 & 24.0  & 2.2 & 20.4 & 5.8 \\ 
220.584 & HNCO 10$_{1,9}$ - 9$_{1,8}$        & 59 & - & - & 11.4 & 4.9 \\ 
220.594 & CH$_3$CN 12$_6$ - 11$_6$           & 326 &  4.7 & 4.7 & 13.1 & 4.0 \\ 
220.641 & CH$_3$CN 12$_5$ - 11$_5$           & 248 & 4.6 & 4.7 & 10.7 & 4.0 \\ 
220.679 & CH$_3$CN 12$_4$ - 11$_4$           & 184 &  5.4 &  4.7 & 16.3 & 4.0 \\ 
220.709 & CH$_3$CN 12$_3$ - 11$_3$           & 134 & 14.6  &  4.7 & 17.2 & 4.0 \\ 
220.730 & CH$_3$CN 12$_2$ - 11$_2$           & 98 & 10.3  & 4.7 & 20.1 & 4.0 \\ 
220.743 & CH$_3$CN 12$_1$ - 11$_1$           & 76 & 10.0  & 4.7 & 22.3 & 4.0 \\ 
220.747 & CH$_3$CN 12$_0$ - 11$_0$           & 69 & 12.3  & 4.7 & 24.1 & 4.0\\ 
229.589 & CH$_3$OH 15$_{4,11}$ - 16$_{3,13}$ E & 374 & 4.2  & 5.3 & 10.0 & 6.0 \\ 
229.759 & CH$_3$OH 8$_{-1,8}$ - 7$_{0,7}$ E & 89 & 12.3  & 6.3 & 22.7 & 4.4 \\ 
229.864 & CH$_3$OH 19$_{5,15}$ - 20$_{4,16}$ A$+$ & 579 & 3.4  & - & 6.0 & 6.0 \\ 
230.027 & CH$_3$OH 3$_{-2,2}$ - 4$_{-1,4}$ E & 39 & 6.0  & 5.8 & 15.8 & 4.0 \\ 
230.538 & $^{12}$CO 2-1$^{b}$ & 17 & 40.0  & 8.6 & 40.0 & 8.6 \\ 
231.061 & OCS 19-18 & 111 & 7.2  & 5.8 & 7.3 & 7.5 \\ \hline 
\end{tabular}
\tablenotetext{a} {We listed lines detected above 5 $\sigma$ level.} 
\tablenotetext{b}{The CO 2-1 spectrum is self-absorbed. The Gaussian fitting is uncertain.} 
\end{center} 
\end{table}

\clearpage

\begin{table}
\caption{Outflow Parameters$^a$} 
\begin{center}
\begin{tabular}{rrlllll} \hline \hline
Outflow & PA($^\circ$) & V$_{Max}^b$ & T$_{dyn}^c$ & Mass & Momentum & Energy \\ 
&      & \kms-1\ & $10^4$ yrs & \msun\ & \msun\kms-1\ & $10^{45}$ ergs \\ \hline 
A & 5  & 50 & 0.5 & 2.5 & 12 & 1.1 \\ 
B & 35 & 25 & 1 & 3.0 & 20 & 2.0 \\ 
C & -60 & 25 & 1 & 3.0 & 20 & 2.0 \\ \hline 
\end{tabular}
\tablenotetext{a}{Parameters are not corrected for orientation of the
outflow.}  
\tablenotetext{b}{The terminal velocity detectable at an rms of 0.09
Jy} 
\tablenotetext{c}{Dynamic timescale computed from $l/V_{Max}$, with $l$
the length of the outflow and $V_{Max}$ the terminal velocity of the outflow.}
\end{center} 
\end{table} 

\clearpage

{\bf Figure Captions}

Figure 1: 225 GHz (1.3mm) and 8.4 GHz (3.6cm) continuum emission toward AFGL 5142. The
1.3 mm continuum peaks (MM-1, MM-2, MM-3, MM-4 and MM-5) are marked by 
stars on both panels.
 The 3.6 cm continuum peaks (CM-1A, CM-1B and CM-2)
are marked by filled squares on the 
right panel. The \h2O maser positions from
\citet{hunter1999} are marked by crosses.  The contour levels are in steps
of 12 mJy/beam for the 225GHz image, and 0.2 mJy/beam  for the 8.4 GHz image.
The synthesized beam is
denoted at the lower-left corner of each panel.

Figure 2:  Line spectra toward MM-1, MM-2, MM-3 and the red-shifted peak
in outflow A (the filled triangle in Figure 3) for the lower side band (LSB) and
upper sideband (USB).  The spectra for the mm 
continuum sources are from
images with a resolution of $1''.3 \times 0''.8$ (The conversion 
factor from Jy/beam to Kelvin is 17 K/Jy).
The spectrum for the outflow position is made with Natural weighting 
and tapering of visibilities, 
resulting in an  angular resolution of $4''.0 \times 3''.3$ (The conversion 
factor from Jy/beam to Kelvin is 1.9 K/Jy).

Figure 3:  Images of the velocity integrated flux of $^{13}$CO J=2-1, 
C$^{18}$O J=2-1, SO $6_5 - 5_4$,
CH$_3$OH $8_{-1,8} -7_{0,7}$ E, OCS J=19-18, and
HNCO $10_{0,10} - 9_{0,9}$ for AFGL
5142.  The range of the integration covers the entire velocity of the line.
The stars mark the position of the mm continuum peaks MM-1, MM-2 and MM-3, MM-4 and MM-5. 
The triangle in the SO
panel marks the position of the spectrum from the outflow A shown in Figure 2.
The contour levels are in steps of
3 Jy~\kms-1~beam$^{-1}$ starting from 3 Jy~\kms-1~beam$^{-1}$,
2 Jy~\kms-1~beam$^{-1}$ starting from 2 Jy~\kms-1~beam$^{-1}$ 
for SO $6_5 - 5_4$, 
and C$^{18}$O J=2-1 lines,  and in steps
of 1 Jy~\kms-1~beam$^{-1}$ starting from 1 Jy~\kms-1~beam$^{-1}$
for CH$_3$OH $8_{-1,8} -7_{0,7}$ E, 
OCS J=19-18 lines, and HNCO $10_{0,10} - 9_{0,9}$ lines.
The size of the primary beam ($57''$) is marked by the dashed circle in the upper-left
panel. The  synthesized beam is
denoted by the shaded ellipse at the lower-left corner of each panel.

Figure 4:  4a: The integrated emission of the CH$_3$CN J=12-11.
The contour levels are in steps of 2 Jy~\kms-1~beam$^{-1}$.
The stars mark the position of the mm continuum peaks MM-1, MM-2, MM-3,
MM-4 and MM-5. The synthesized beam is denoted by the shaded ellipse
at the lower-left corner.
4b: The observed (thin lines) and model (thick lines) spectra toward the position 
of MM-1 and MM-2.

Figure 5:  $^{12}$CO 2-1 emission toward AFGL 5142.  
The stars mark the position of the mm continuum peaks MM-1, MM-2, MM-3,
MM-4 and MM-5.  
The plus symbols denote the near infrared \h2
emission knots \citep{hunter1999}.  
Arrows mark the three molecular outflows A, B and C.  The
corresponding velocity range of the $^{12}$CO emission is labeled at the
upper-right corner of each panel.  The synthesized beam is denoted by the shaded ellipse
at the lower-left corner of the first panel.  
5a:  High velocity $^{12}$CO 2-1 emission from the SMA data alone. 
The contour levels are in steps of 0.3 mJy/beam. 
5b:  Low velocity $^{12}$CO 2-1 emission from combined SMA and CSO data.
The contour levels are in steps of 0.6 mJy/beam.

Figure 6:  The position velocity diagram of the $^{12}$CO 2-1 emission.  The
CO data are from SMA alone.  The cut is along the axis of outflows A, B and C,
with position offset at MM-2.  The contour levels are in steps of 0.4 Jy/beam.
The vertical dashed lines mark the cloud systemic velocity of -3.9 \kms-1.

Figure 7:  Channel maps of the SO $6_5 - 5_4$ line.  The contour levels are in steps
of 0.4 Jy/beam.  The stars mark the position of the mm continuum peaks MM-1, MM-2, MM-3,
MM-4 and MM-5.  The plus symbols denote the near infrared \h2
emission knots \citep{hunter1999}. Arrows
mark the three molecular outflows A, B and C. The  synthesized beam is
denoted by the shaded ellipse at the lower-left corner of the first panel.

Figure 8:  Channel maps of the CH$_3$OH $8_{-1,8} -7_{0,7}$ E line.  
The contour levels are in steps of 0.4 Jy/beam.  
The stars mark the position of the mm continuum peaks MM-1, MM-2, MM-3,
MM-4 and MM-5. The plus symbols denote the near infrared \h2
emission knots \citep{hunter1999}. 
Arrows mark the three molecular outflows A, B and C. The  synthesized beam is
denoted by the shaded ellipse at the lower-left corner of the first panel.

Figure 9:  The SiO 2-1 emission from \citet{hunter1999} in contours
overlaid on the $^{12}$CO 2-1 emission in greyscale for outflow A.  The cut of
the position velocity plot is along the outflow axis with position offset from
MM-2. The SiO emission is plotted in steps of 18 mJy~beam$^{-1}$.

Figure 10:  The ratios of the \nh3 (J,K)=(3,3) over (1,1), and
\nh3 (J,K)=(4,4) over (1,1) emission (color)  overlaid on the
integrated emission of the \nh3 (J,K)=(1,1) line (contours).  The images are
taken from \citet{zhang2002}.  The higher ratios denote higher
temperatures.  Arrows mark the three molecular outflows A, B and C.
The stars mark the position of the mm continuum peaks MM-1, MM-2, MM-3,
MM-4 and MM-5.

\clearpage

\begin{figure} 
\plotone{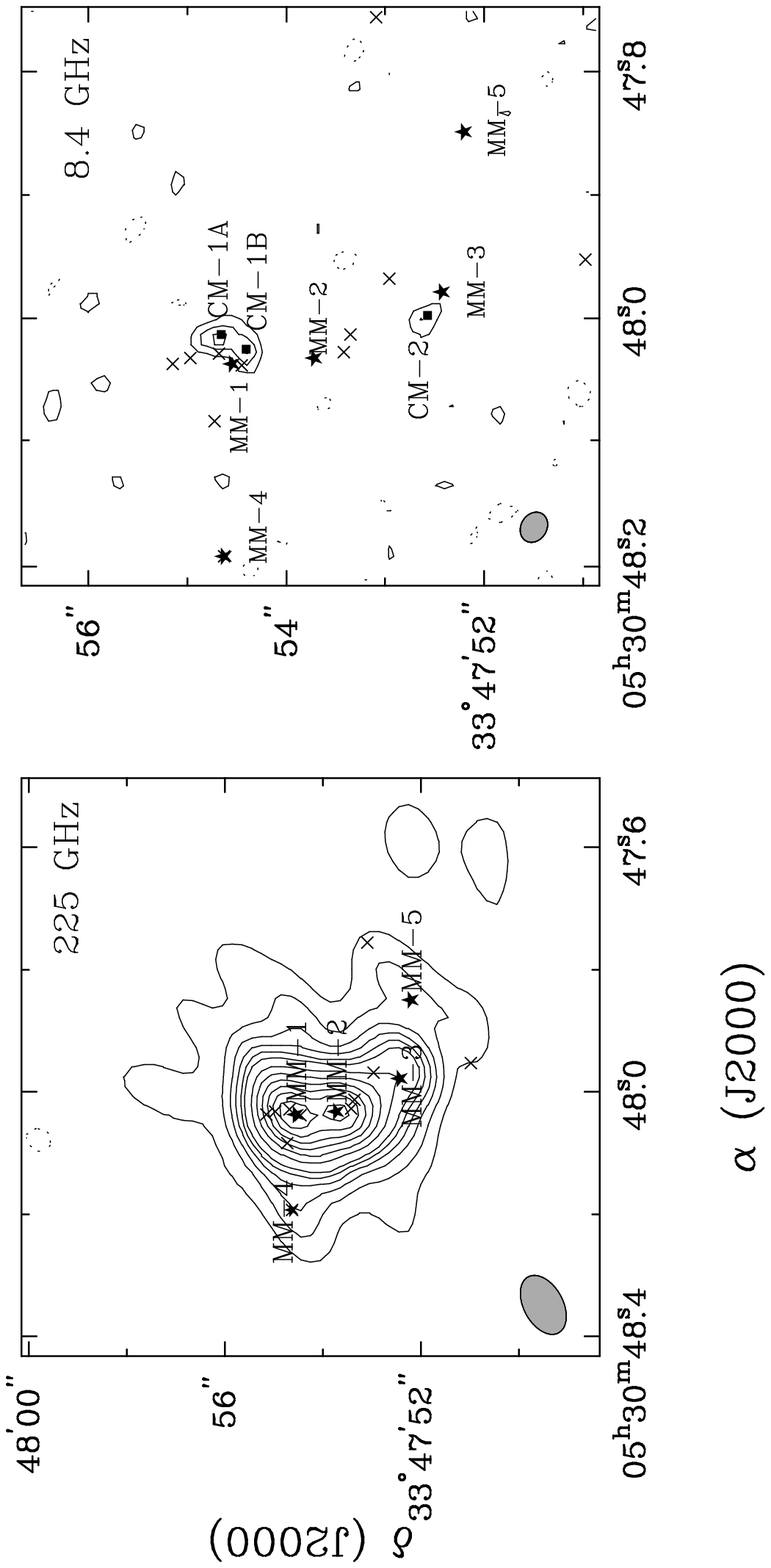}
\caption{}  
\end{figure}

\begin{figure} 
\figurenum{2a}
\plotone{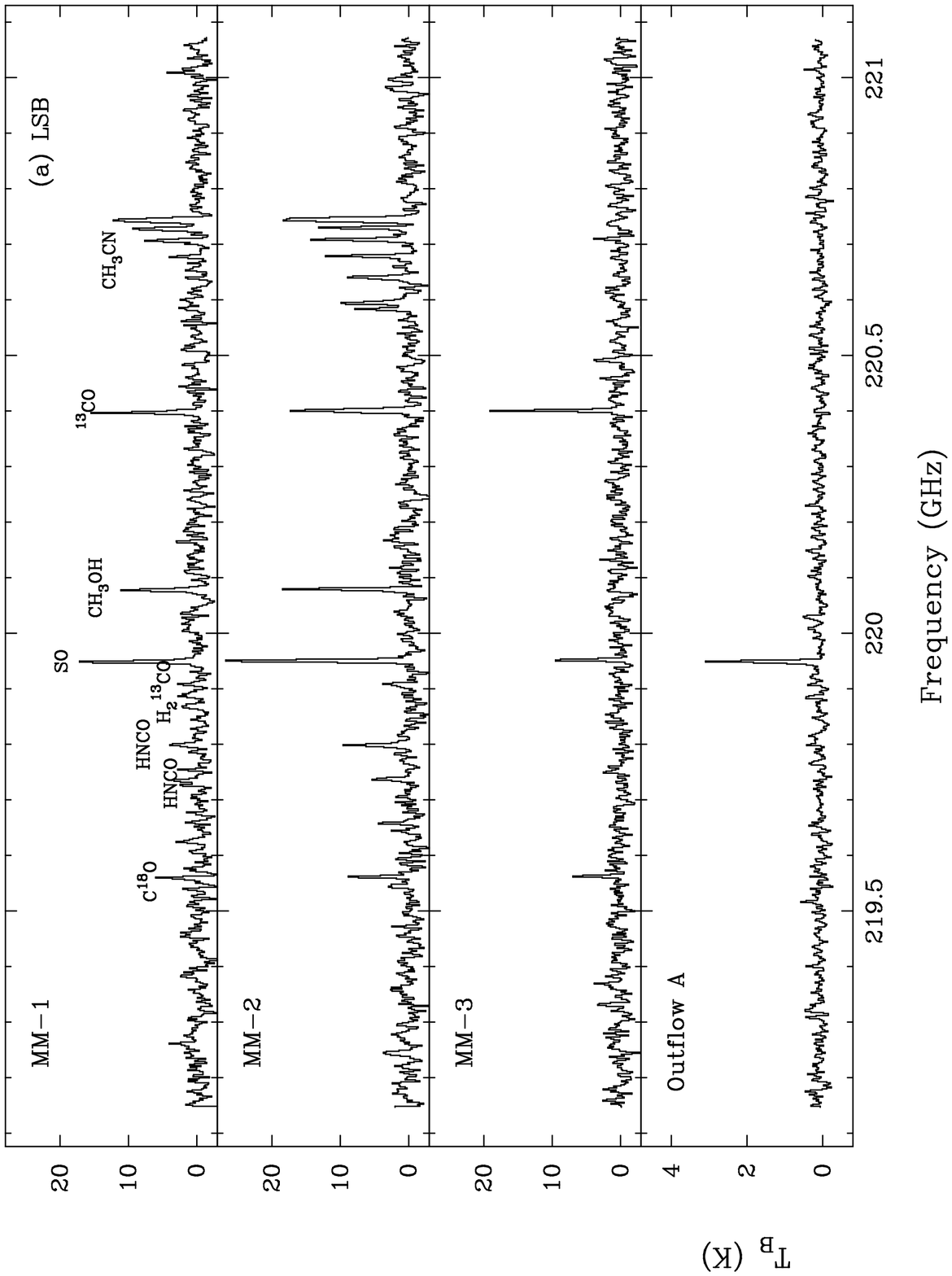}
\caption{}
\end{figure}

\begin{figure} 
\figurenum{2b}
\plotone{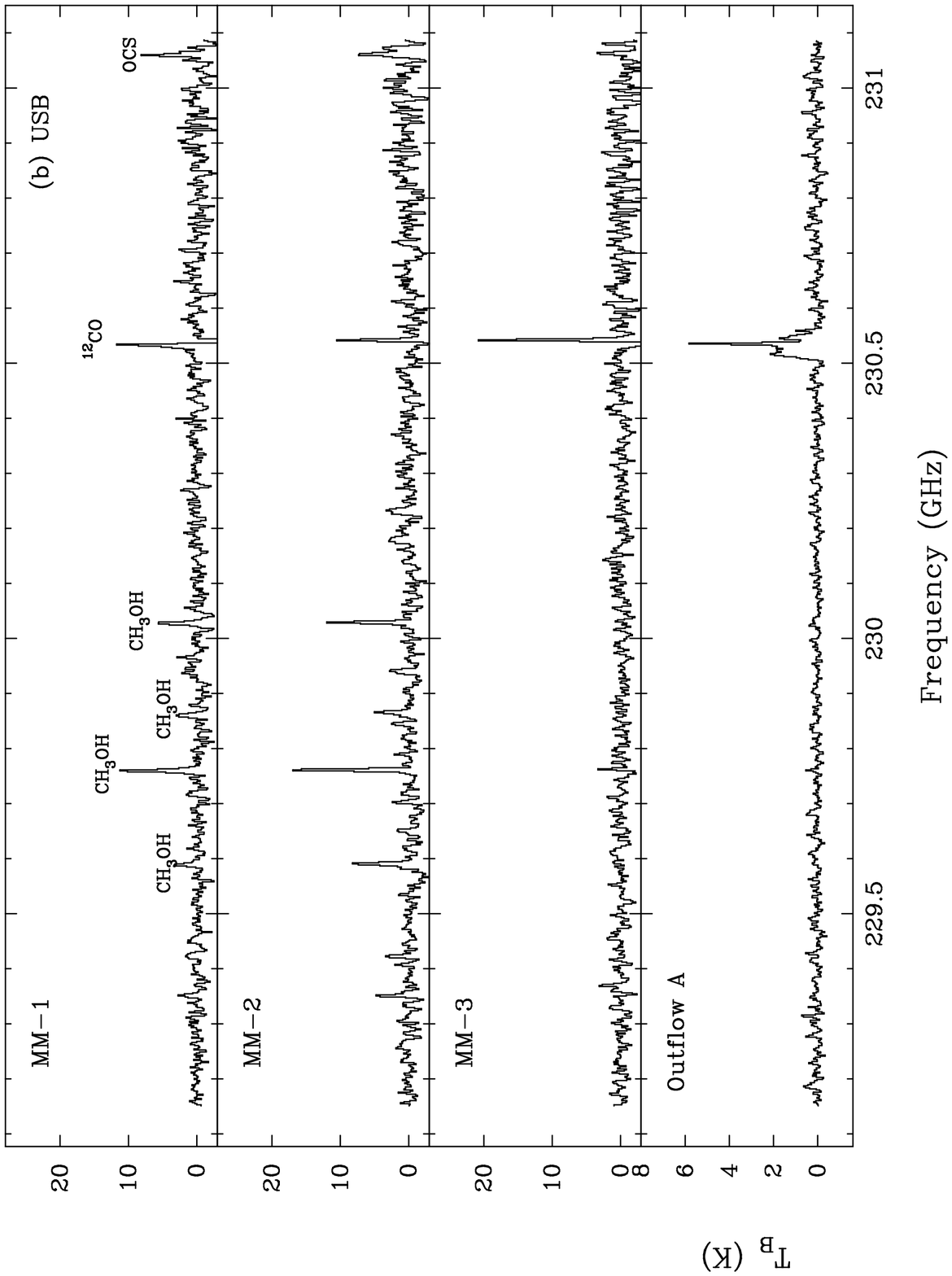}
\caption{}
\end{figure}

\begin{figure} 
\figurenum{3}
\plotone{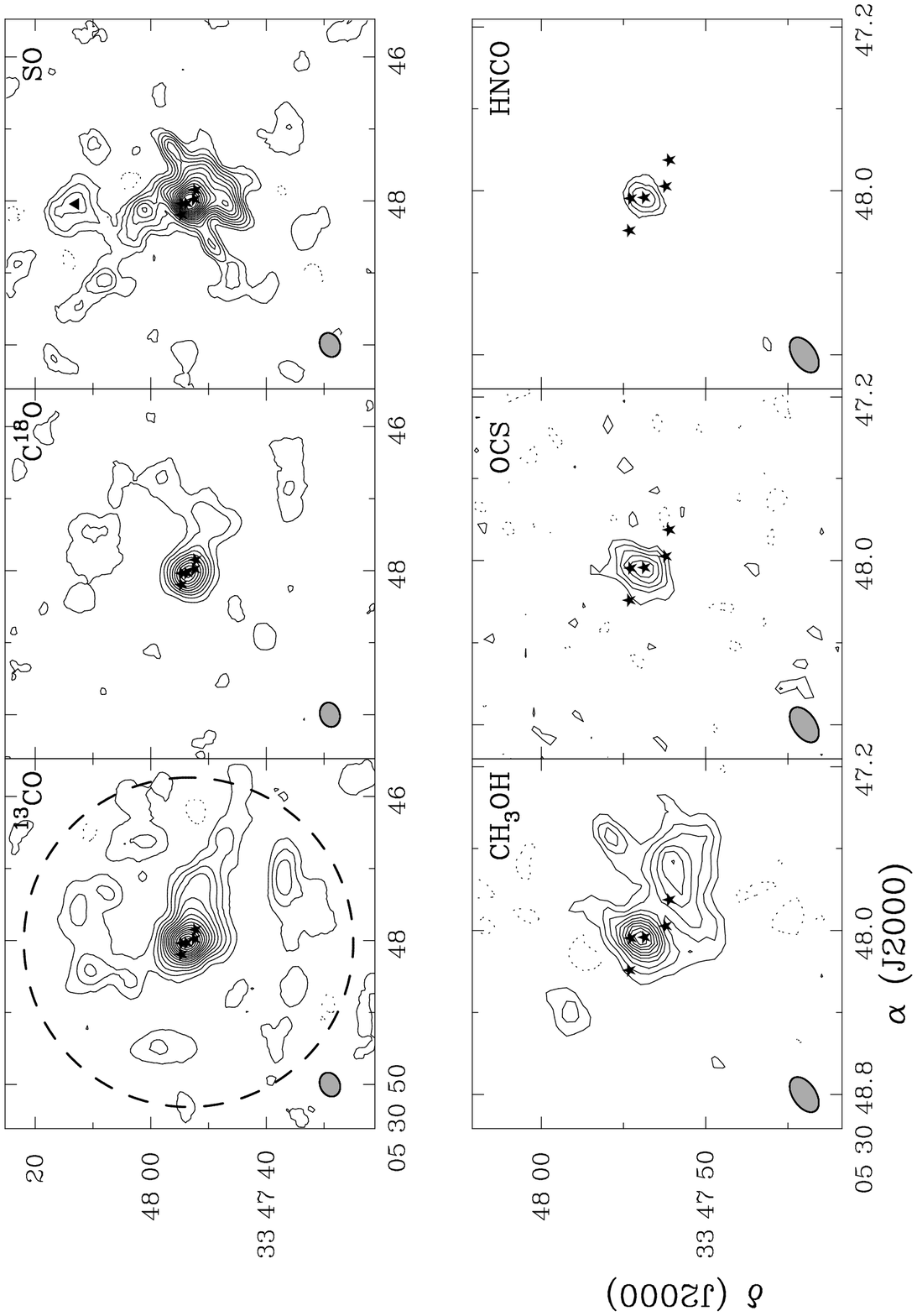}
\caption{}
\end{figure}

\begin{figure} 
\figurenum{4a}
\plotone{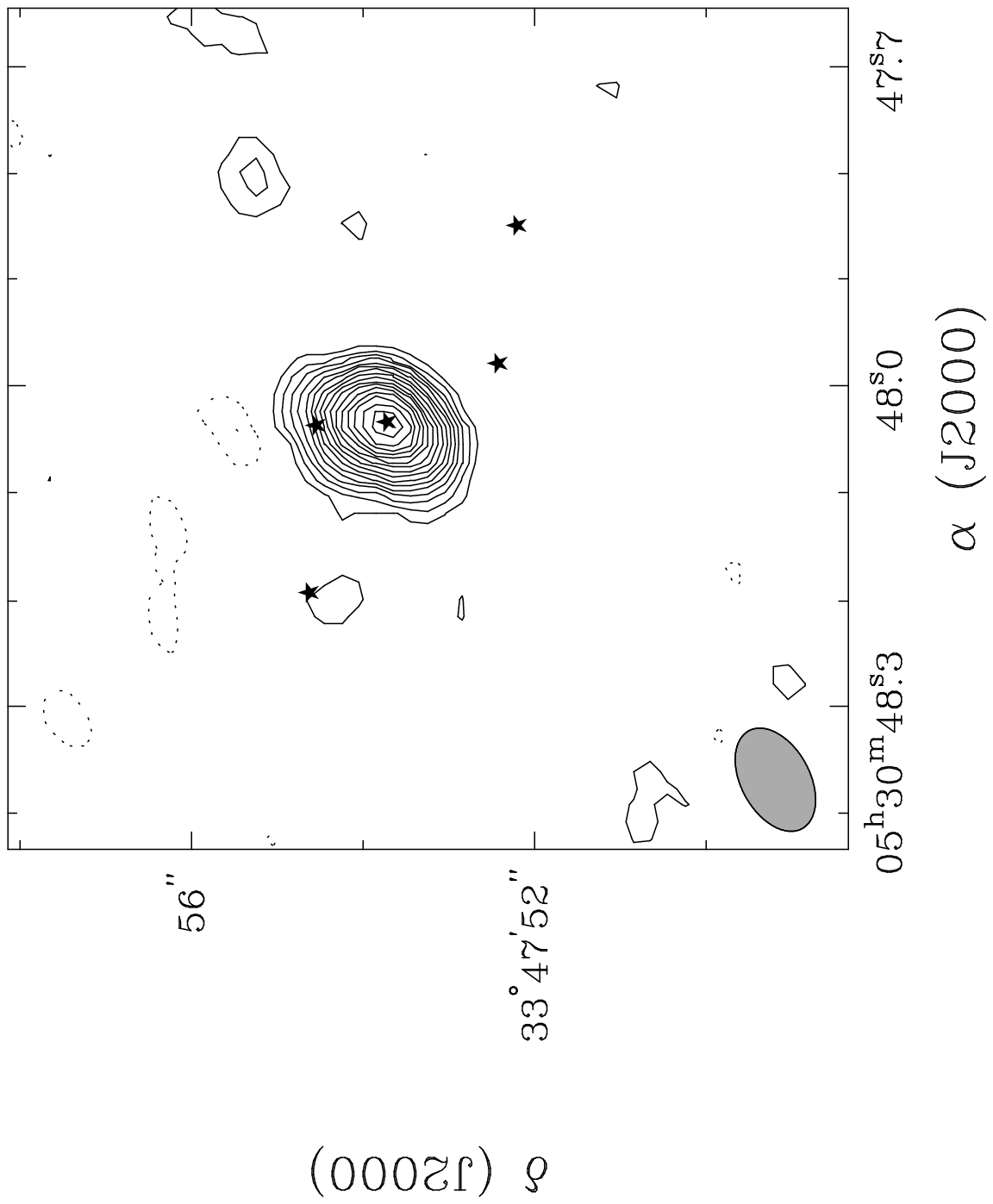}
\caption{} 
\end{figure}

\begin{figure} 
\figurenum{4b}
\plotone{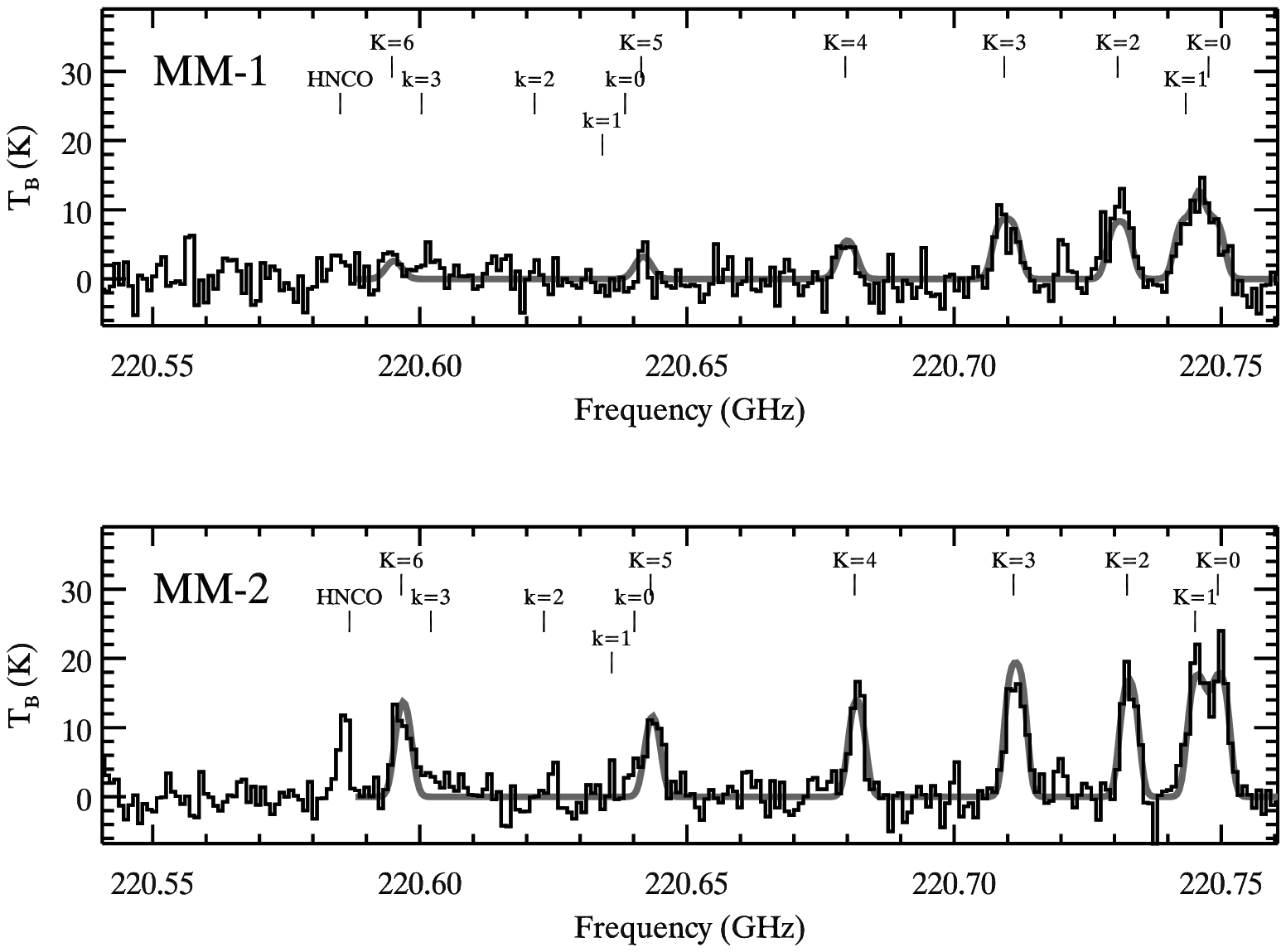}
\caption{} 
\end{figure}

\begin{figure}
\figurenum{5a}
\plotone{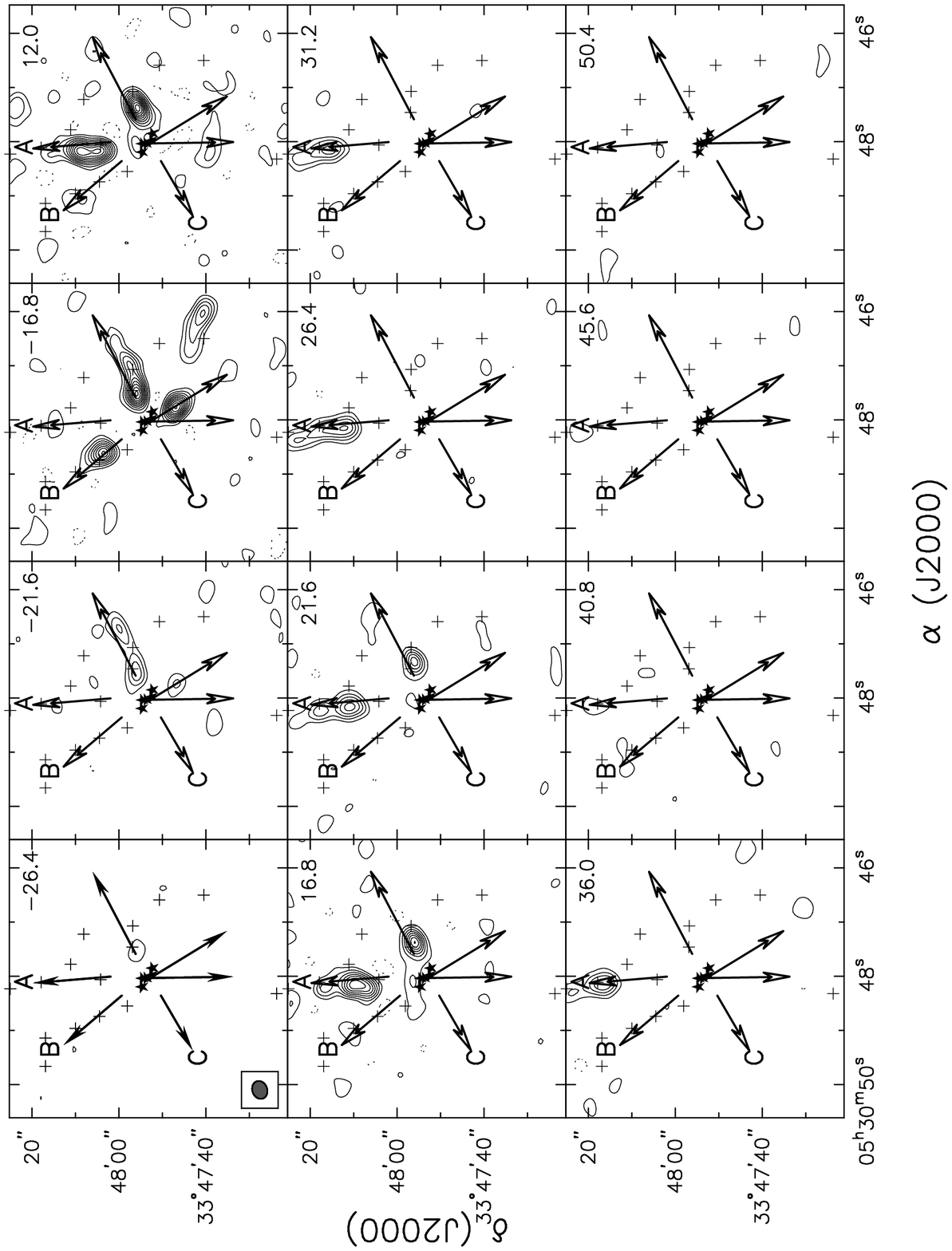} 
\caption{}
\end{figure}

\begin{figure}
\figurenum{5b}
\plotone{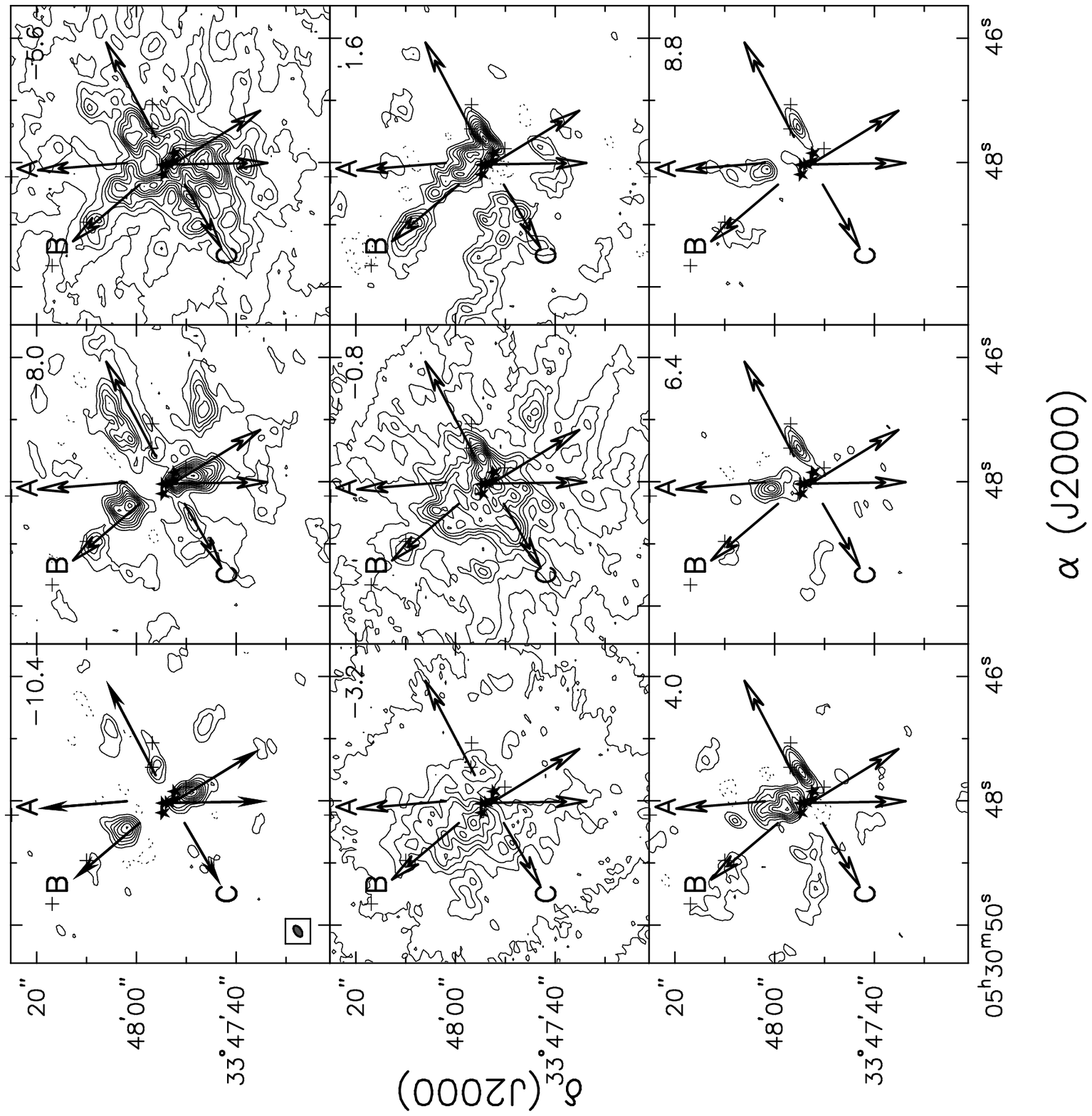} 
\caption{}
\end{figure}

\begin{figure}
\figurenum{6}
\plotone{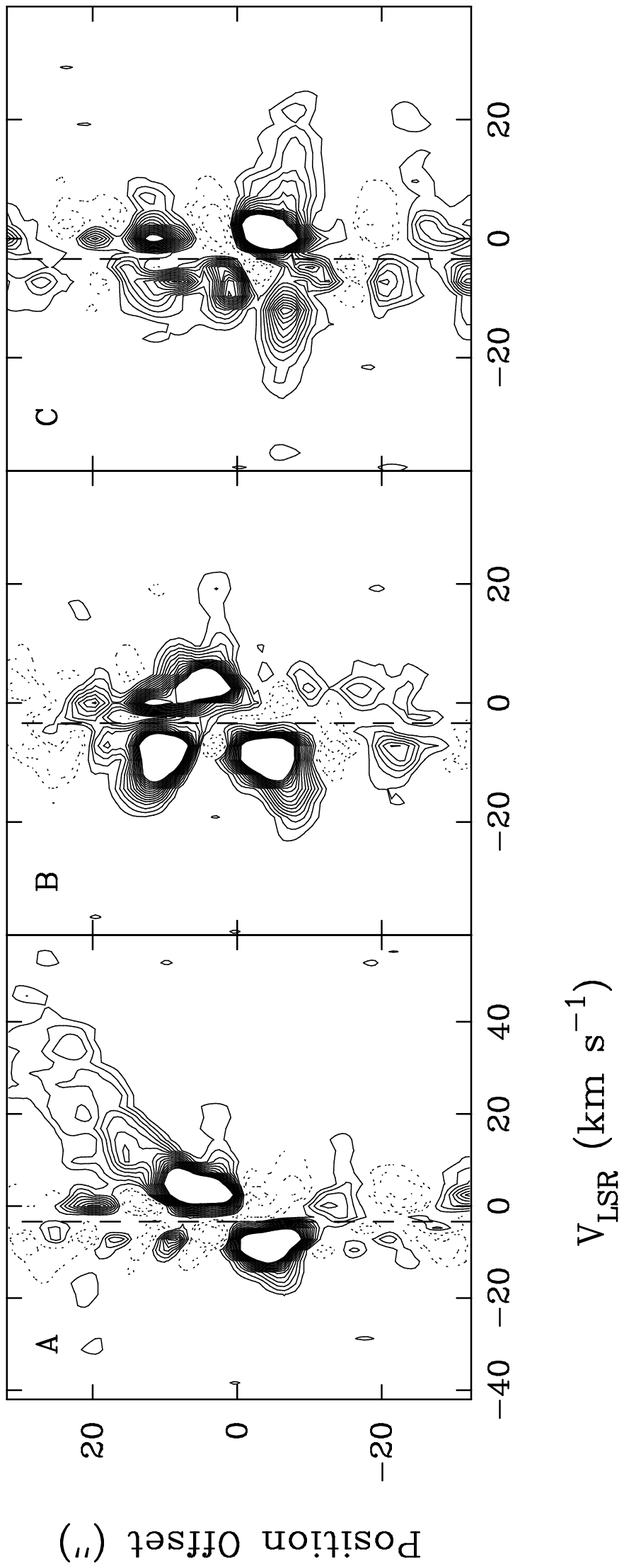}
\caption{}
\end{figure}

\begin{figure} 
\figurenum{7}
\plotone{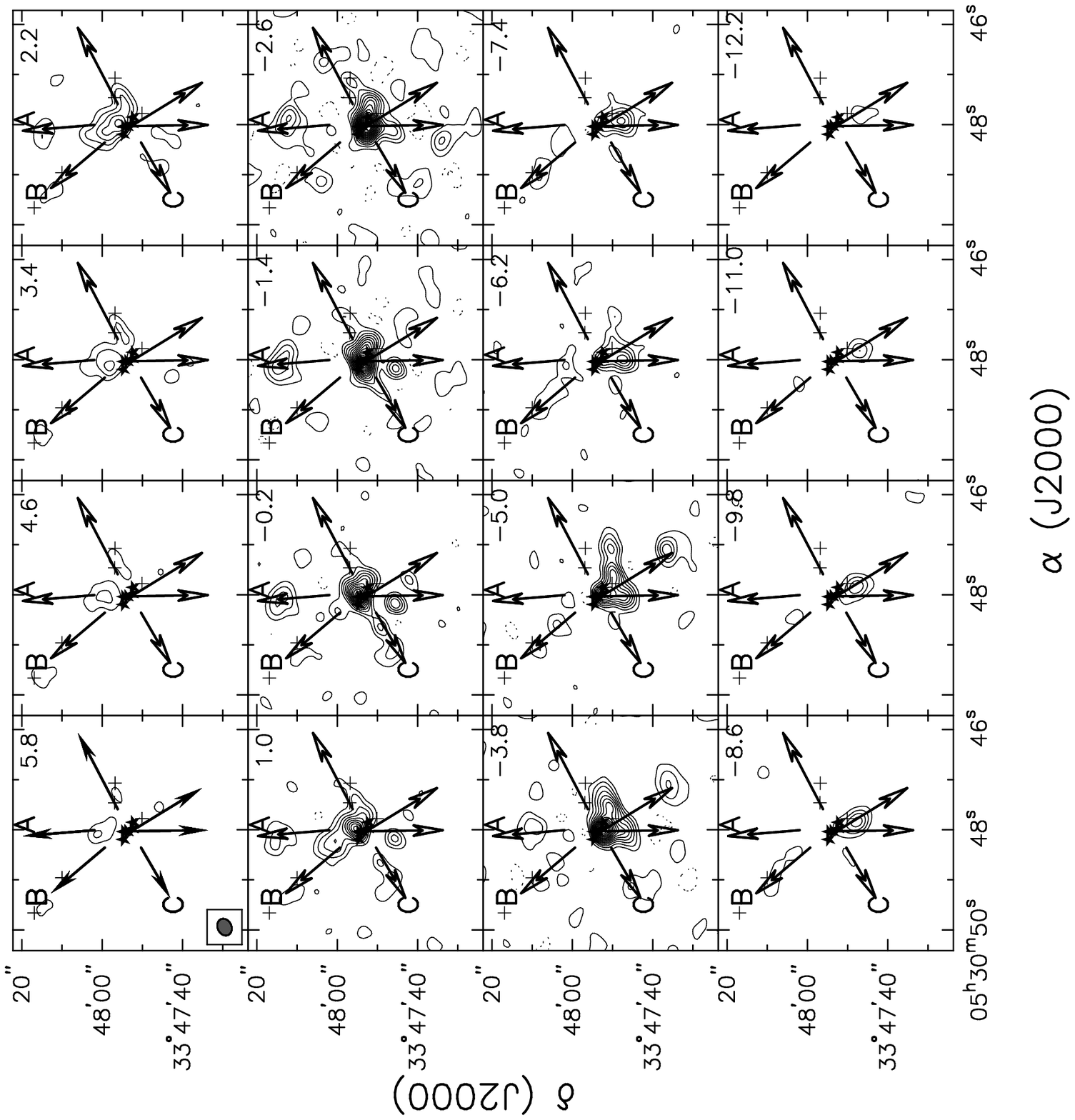}
\caption{}
\end{figure}

\begin{figure} 
\figurenum{8}
\plotone{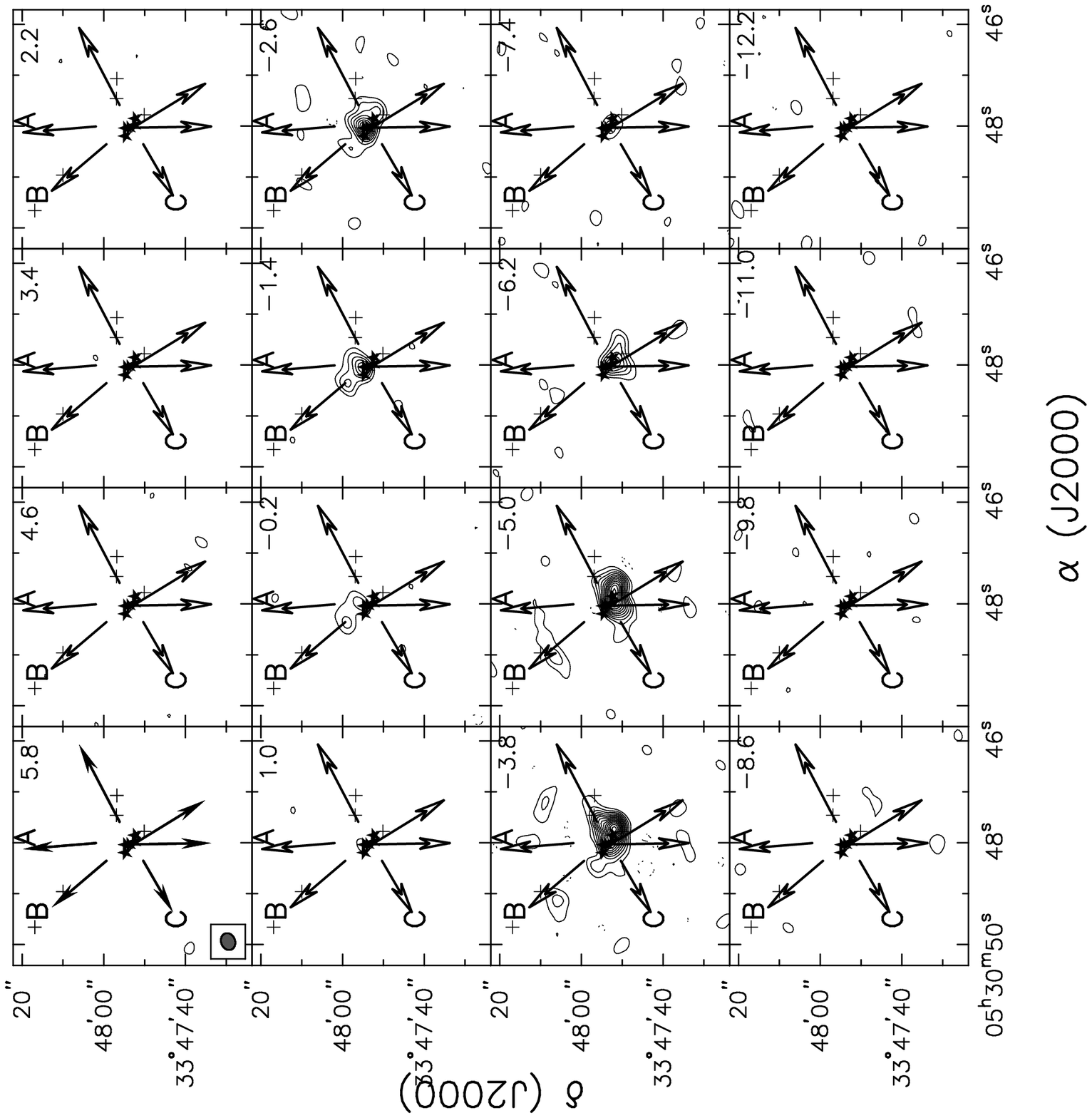}
\caption{}
\end{figure}

\begin{figure} 
\figurenum{9}
\plotone{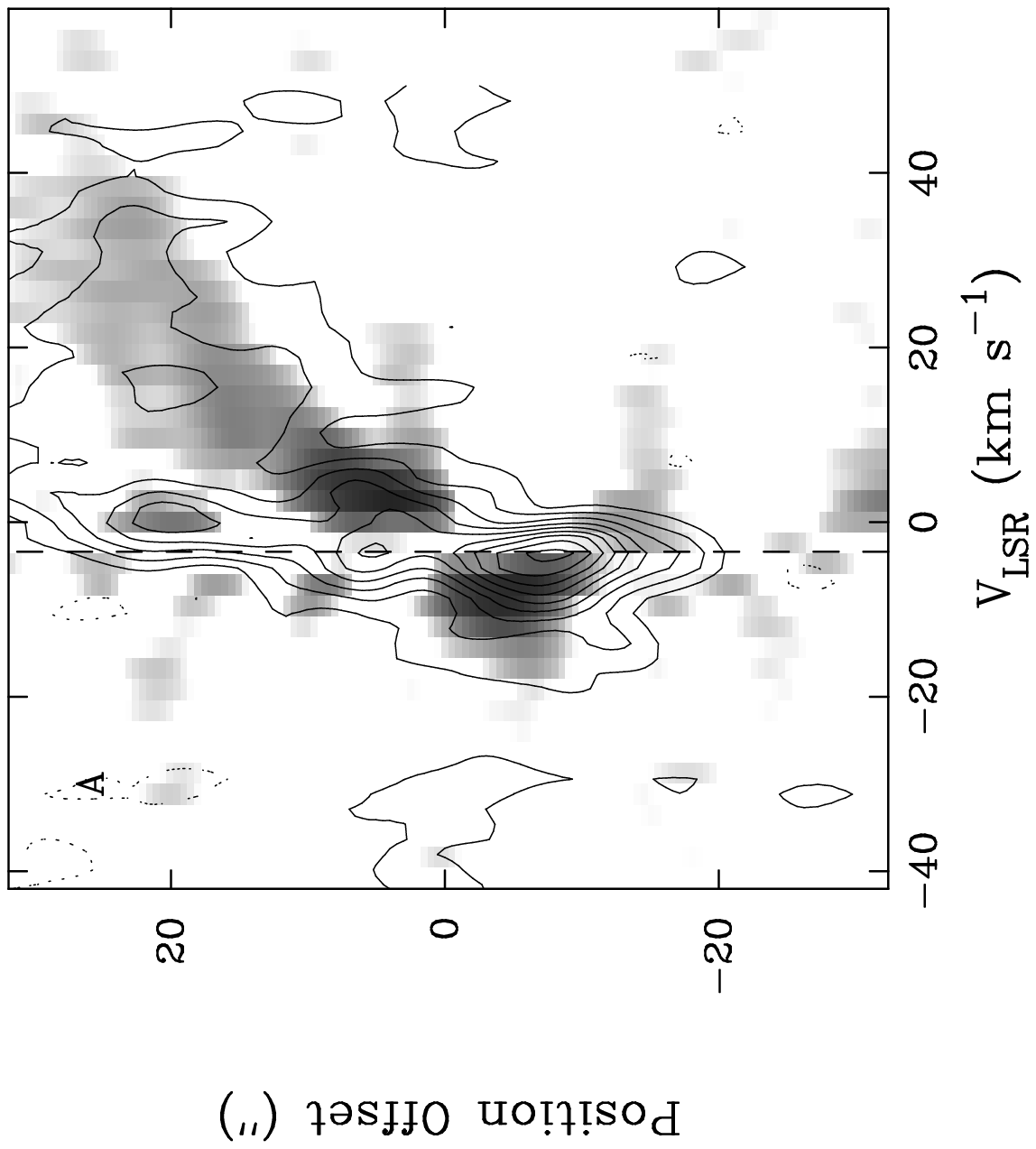}
\caption{}
\end{figure}

\begin{figure}
\figurenum{10}
\plotone{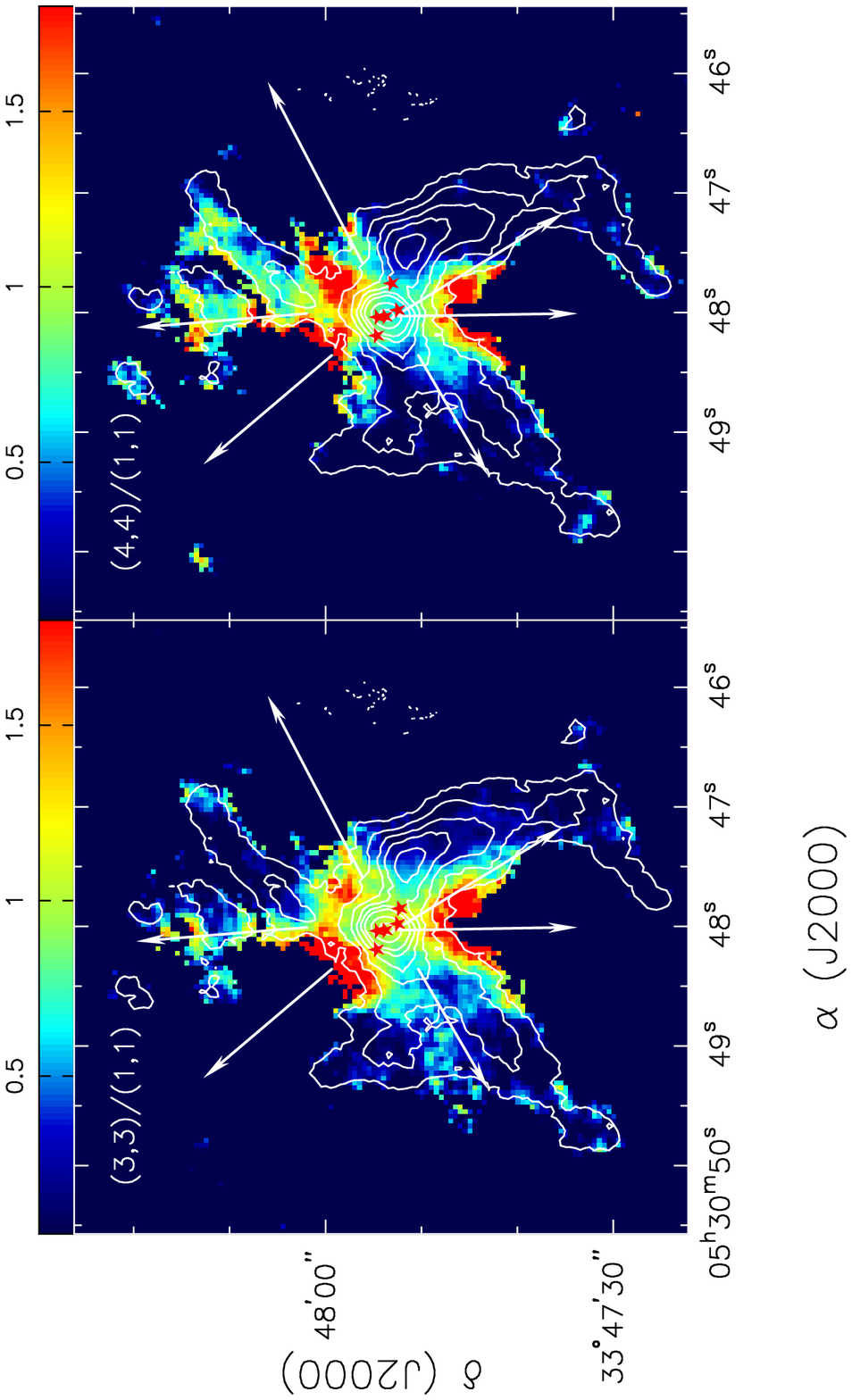}
\caption{} 
\end{figure}

\end{document}